\documentclass{aa}  
\usepackage{caption}
\usepackage{subcaption}

\usepackage{float}
\usepackage{blindtext}

\usepackage{multicol}
\usepackage{graphicx}
\usepackage{tabularx}
\usepackage{rotating}
\usepackage{longtable}
\usepackage{placeins}
\usepackage{afterpage}
\usepackage{pdflscape}
\usepackage[table,xcdraw]{xcolor}
\usepackage{balance}
\usepackage{adjustbox}
\usepackage{orcidlink}
\usepackage{txfonts}

\begin{document}

   \title{Enhanced detection limits in the SHINE F150 survey through the regime switching model}
   \subtitle{Optimizing thresholds and investigating environmental noise}

   \author{M. Sabalbal\orcidlink{0009-0005-9576-5229}\inst{1} \and
           O. Absil\orcidlink{0000-0002-4006-6237} \inst{1}\fnmsep\thanks{F.R.S.-FNRS Research Director}\and
           C.-H. Dahlqvist\orcidlink{0000-0003-4994-9244} \inst{1}\and
          P. Delorme \orcidlink{0000-0002-2279-410X}\inst{2}
          }
    
   \institute{
    STAR Institute, Université de Liège, Allée du Six Août 19C, 4000, Liège, Belgium\\
    \email{mariam.sabalbal@uliege.be}
    \and
         IPAG, Univ Grenoble Alpes, CNRS, Grenoble, France
         }
   \date{Received 18 April 2025 / Accepted 4 December 2025}

  \abstract
    {In high-contrast imaging, a novel detection algorithm for angular differential imaging (ADI) sequences has recently been introduced: the regime switching model (RSM). This advanced statistical tool enhances the distinction between planetary signals and bright speckles by simultaneously combining multiple ADI-based post-processing techniques.

    }
    {In this study, we apply the RSM algorithm to analyze the F150 sample from the SHINE high-contrast imaging survey carried out with VLT/SPHERE, aiming to enhance detection limits and identify new exoplanet candidates. Additionally, we investigate how environmental conditions influence post-processed noise distributions and detection thresholds.
    }
    {We generated detection maps and contrast curves for 213 observations in the F150 SHINE sample using the RSM algorithm. A clustering approach based on environmental parameters was used to group observations with similar noise characteristics. We propose two methods for defining radial detection thresholds in the RSM maps: fitting a lognormal distribution to the post-processed noise and maximizing the F1 score. We also assessed the performance of various combinations of post-processing techniques within the RSM framework to identify optimal configurations.}
    {This study demonstrates the utility of clustering based on observational parameters, effectively distinguishing features such as wind-driven halos and low-wind effects. Detection thresholds vary significantly across clusters, differing by up to a factor of ten, highlighting the importance of considering observational environments. Lognormal thresholds provide conservative, noise-aware limits, while F1 score-based thresholds offer observation-specific results, with both showing compatibility overall. RSM improves detection limits by an average factor of two at $1\arcsec$ and five at inner working angles compared to standard principal component analysis processing. This study reports more than 30 newly detected signals, including one promising candidate awaiting second-epoch confirmation.
    }
    

   \maketitle

\section{Introduction}
The field of high-contrast imaging (HCI) has undergone remarkable advancements in recent years, driven by innovations in coronagraphs, adaptive optics, sophisticated observing strategies, and advanced post-processing techniques. These developments have made direct imaging a powerful tool for detecting hot, massive Jupiter-such as exoplanets at wide orbital separations, capable of capturing planets that are up to $10^6$ times dimmer than their host stars in the infrared. This capability has led to the detection and characterization of dozens of young exoplanets, offering invaluable insights into planetary formation processes \citep[see][for a recent review]{2023ASPC..534..799C}. Pushing detection limits further to identify fainter, closer-in planets would bridge the gap between direct imaging and indirect methods, providing richer information about planetary formation pathways, enabling detailed atmospheric characterization of a larger sample of exoplanets, and expanding our understanding of exoplanet demographics.

The Spectro-Polarimetric High-contrast Exoplanet Research \citep[SPHERE;][]{2019A&A...631A.155B} instrument is a second-generation extreme adaptive optics system at the Very Large Telescope (VLT), equipped with advanced coronagraphs feeding imaging and spectroscopic cameras. SPHERE has been central to several exoplanet imaging campaigns, including the SPHERE High-contrast Imaging survey for Exoplanets \citep[SHINE;][]{2017sf2a.conf..331C}. SHINE targets around 400 young, nearby, and relatively bright stars 
to detect and characterize giant exoplanets, contributing to our understanding of their architectures and formation \citep[see][for further details]{2021A&A...651A..70D}. The survey utilizes both of SPHERE’s near-infrared imaging cameras -- the dual band imager IRDIS \citep[][]{2008SPIE.7014E..3LD} and the integral field spectrograph  \citep[IFS;][]{2010SPIE.7735E..0VC} --, employing angular differential imaging \citep[ADI;][]{2006ApJ...641..556M} for high-contrast observations. The SHINE survey has enabled key discoveries, such as HIP~65426b \citep{2017A&A...605L...9C} and PDS~70b \citep{2018A&A...617L...2M}; has refined constraints on giant planet occurrence rates \citep{2021A&A...651A..71L} and demographics of young exoplanets within 300~AU \citep{2021A&A...651A..72V}; and has spatially resolved several circumstellar disks \citep{2016A&A...586L...8L,2017A&A...601A...7F,2016A&A...595A.114D}. In their analysis of a 150-star subset from the SHINE survey, referred to as the F150 sample, \citet{2021A&A...651A..71L} evaluated the survey’s detection capabilities. Utilizing SPHERE’s IRDIS H-band mode, the median detection performance achieved a $5 \sigma$ contrasts of 11.8 mag at 200 mas and 13.1 mag at 800 mas. These results were obtained by applying standard post-processing techniques, including principal component analysis \citep[PCA;][]{Soummer_2012ApJ...755L..28S,2012MNRAS.427..948A} and the template-based locally optimized combination of images \citep[TLOCI;][]{2014SPIE.9148E..0UM}.

In recent years, more advanced post-processing algorithms such as ANDROMEDA \citep{2015A&A...582A..89C}, PACO \citep{2018A&A...618A.138F,2020A&A...637A...9F,2024MNRAS.527.1534F}, SODINN~/~NA-SODINN \citep{2018A&A...613A..71G,2023A&A...680A..86C}, regime switching model \citep[RSM;][]{2020A&A...633A..95D,2021A&A...646A..49D,2021A&A...656A..54D}, and 4S \citep{2025AJ....169..194B} have been developed, with the goal of optimizing noise modeling and enhancing planetary signal detection -- particularly for faint companions at close separations. Notably, \citet{2023A&A...675A.205C} applied PACO to the SHINE F150 sample, improving the median contrast limits by a factor of five at inner working angles and a factor of two overall, and demonstrating the impact of more refined statistical modeling.

Building on this progress, we revisit the SHINE F150 sample using the RSM algorithm, which has shown excellent performance in the Exoplanet Imaging Data Challenge \citep[EIDC;][]{10.1117/12.2574803} by achieving high F1 scores and a low false positive rate. Unlike spatial-only modeling approaches, RSM captures both the spatial and temporal evolution of pixel intensities across residual, de-rotated frames using a Markov regime-switching framework applied to outputs from multiple PSF subtraction techniques. This enables it to fully exploit the temporal structure of high-contrast imaging sequences, offering improved sensitivity and robustness in exoplanet detection. In this study, we define detection thresholds tailored to RSM maps while accounting for noise variations across observing conditions. We compare two thresholding strategies, one assuming a lognormal noise distribution, and another based on maximizing the F1 score, and we assess their relative performance. We further evaluate how different post-processing algorithm combinations affect RSM sensitivity and present updated detection limits for the SHINE survey based on this reanalysis.

The structure of this paper is as follows: Sect.~\ref{section_survey_description} outlines the F150 sample selection criteria. In Sect.~\ref{section_environmental_conditions} we describe the observational conditions in the dataset. Section~\ref{section_rsm_algorithm} presents an explanation of the RSM algorithm and its application in this study. In Sect.~\ref{section_noise_behavior_in_RSM} we examine two methods for setting detection thresholds and their atmospheric context. In Sect.~\ref{section_contrast_curves} we present contrast curves for different detection thresholds and post-processing combinations, highlighting the optimal approach within the RSM framework, and we discuss the improvements in the SHINE detection limits.
In Sect.~\ref{section_identification_point_sources} we present the point sources detected by RSM in the F150 sample and compare them with those identified by other algorithms, such as PACO, and we include a discussion of newly detected candidates.

\section{The data sample}\label{section_survey_description}

In this study, we use the pre-reduced F150 sample from the SHINE survey, observed in H23 bands with ADI mode, covering an approximately $9\arcsec \times 9\arcsec$ field of view. The High Contrast Data Center \citep[HC-DC,][]{2017sf2a.conf..347D,2018A&A...615A..92G,2019A&A...631A.155B,2016SPIE.9908E..34M} pre-processed these observations, resulting in a dataset of 343 observations. Atmospheric conditions during these observations varied significantly, with seeing values ranging from $0\farcs4$ to $3\arcsec$ and Strehl ratios from 0.1 to 0.95. To ensure data quality and consistency, a preselection of datasets was necessary based on these atmospheric parameters.

The preliminary selection of data was based on several parameters, including seeing, Strehl ratio, raw contrast, number of frames, full width at half maximum (FWHM), and an assessment of the features in the PSF and the science cubes. Information such as seeing, wind speed and coherence time were primarily derived from differential image motion monitor \citep[DIMM,][]{1990A&A...227..294S} measurements, as recommended by \citet{2017AO4ELT5}, while the Strehl ratio was empirically measured from the PSF. Additional parameters, such as raw contrasts for science cubes and frame quality assessments, were provided by the HC-DC reduction pipeline. 
In cases where DIMM seeing measurements were unavailable, they were inferred from the available seeing values provided by the SPHERE real-time computer SPARTA \citep{2006SPIE.6272E..10F} using the relationship between DIMM and SPARTA seeing established in \citet{2017AO4ELT5}.

To ensure a sample of stars with observing conditions ranging from fair to excellent, observations that met any of the following conditions were also discarded: seeing larger than $2\arcsec$, Strehl ratio less than 0.5, FWHM greater than 5.5~pixels, number of frames in the ADI sequence fewer than 40, and non-physical raw contrasts at 500 mas (negative or greater than one). In case of multiple observations of the same star within five consecutive nights, only the observation with the best conditions was retained. 
Finally, we assessed the quality of both coronagraphic and non-coronagraphic cubes using a combination of tools: the HC-DC frame quality assessment, correlation functions from the VIP package \citep{2017AJ....154....7G,2023JOSS....8.4774C}, and visual inspection. Cubes containing a majority of corrupted frames were discarded. In addition, science cubes in which the star was located outside the coronagraph mask, or where the off-axis PSF showed strong secondary lobes caused by the low-wind effect \citep[see][]{2018SPIE10703E..2AM}, were also excluded.

These filtering steps resulted in a total of 213 observations for 150 stars. These constitute the data sample used in this paper. Additional information about the list of targets and their observing conditions is provided in Section~\ref{data_availability}.

\section{Clustering based on environmental conditions} \label{section_environmental_conditions}

The quality of observations and specific features within coronagraphic images can be characterized by several environmental and instrumental parameters. Factors such as seeing, wind speed, coherence time, Strehl ratio, and raw contrast collectively indicate observation quality and provide insights into SPHERE-specific features. For instance, the wind-driven halo \citep{2020A&A...638A..98C} arises under high wind velocity and short coherence times, while the low-wind effect \citep[LWE;][]{2018SPIE10703E..2AM} appears at low wind speeds. These features can severely degrade SPHERE image quality, reducing the achievable contrast levels \citep[see] [for more details]{2019Msngr.176...25C}.

\begin{figure}[t]
    \centering
    \includegraphics[width=0.5\textwidth]{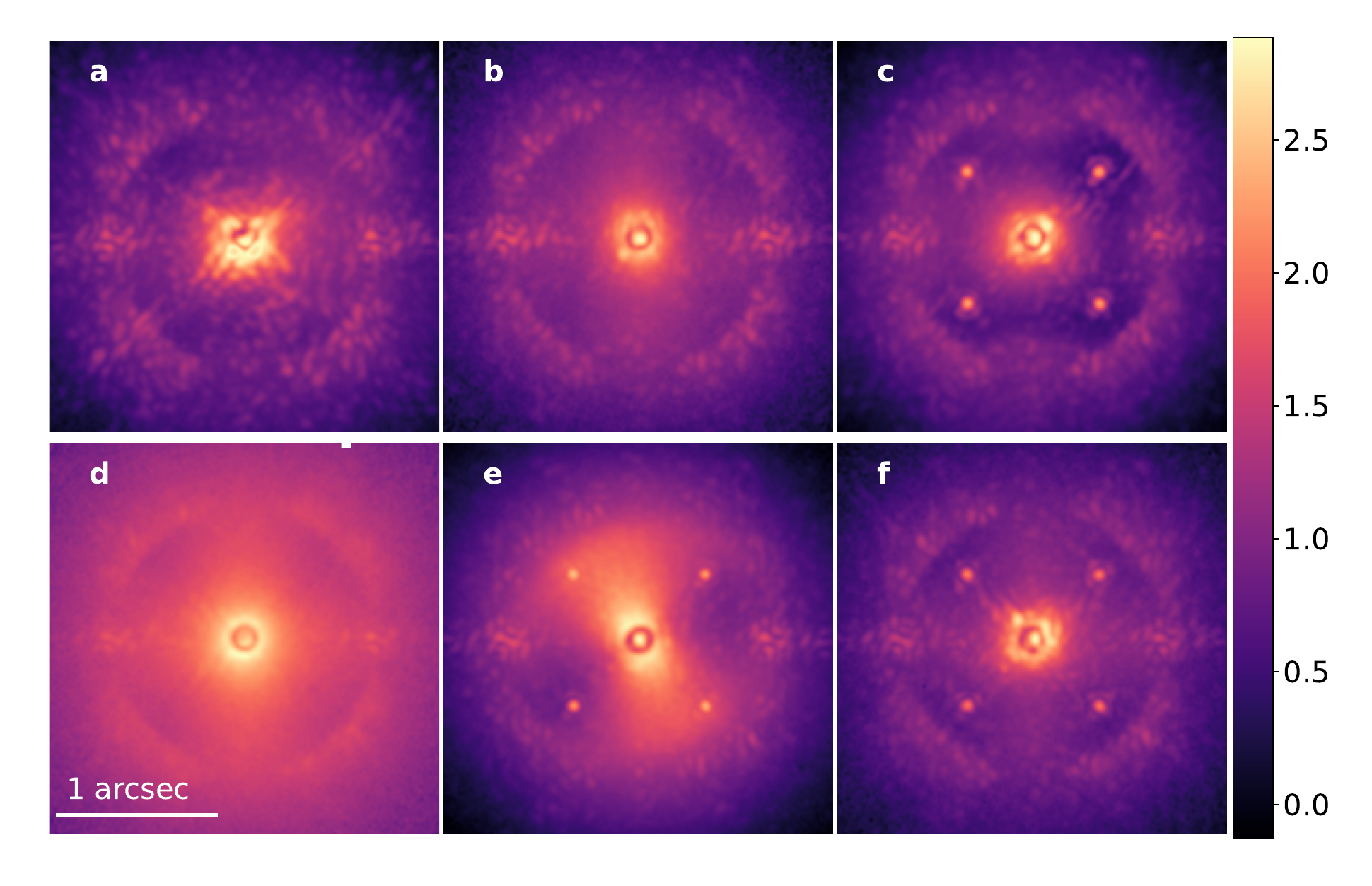}
    \caption{Images of single frames from various observations across different clusters. The images are labeled from (a) to (f), which correspond to clusters 1 to 6 presented in log scale. 
    }
    \label{median_frames}
\end{figure}

We adopted the clustering method outlined in \citet{2022A&A...666A..33D}, grouping observations with similar characteristics using the k-means clustering algorithm from the \texttt{scikit-learn} package. This approach not only helps identify distinct classes of SPHERE observations but also reduces computational load when optimizing PSF subtraction techniques and RSM parameters (as described in Section \ref{section_rsm_algorithm}). The parameters selected for clustering include the number of frames in the ADI sequence, seeing, wind speed, parallactic angle range, Strehl ratio, raw contrast at 500 mas, and coherence time. Observations were grouped into six clusters. The number of clusters was determined by maximizing the silhouette score, a metric used to evaluate the quality of clustering. For each cluster, k-means automatically selected a representative observation, i.e., the dataset closest to the cluster centroid. In the following, the terms cluster center and representative observation are used interchangeably. The parameters for PSF subtraction and the RSM algorithm could then be optimized for these representative datasets, and generalized to other observations within the same cluster. To ensure unbiased parameter selection, we made sure that none of the cluster centers contained datasets with bright astrophysical signals. A pixel correlation test on science cubes was used to validate the effectiveness of clustering in capturing noise features (see Appendix~\ref{appendix_correlations}), and it supports the use of the k-means elected centers for parameter optimization. Figure~\ref{median_frames} displays a $200 \times 200$ pixel crop of representative individual frames from different clusters. These images illustrate the varying noise behavior across clusters, with some frames showing secondary lobes at close separations from the star indicative of the low-wind effect (Fig.~\ref{median_frames}a), while others exhibit strong elongated (i.e. butterfly) patterns characteristic of the wind driven halo (Fig.~\ref{median_frames}e).

\begin{figure*}[htp]
    
    \begin{center}
    
        \begin{subfigure}[b]{.46\textwidth}
            \centering
            \includegraphics[width=\textwidth]{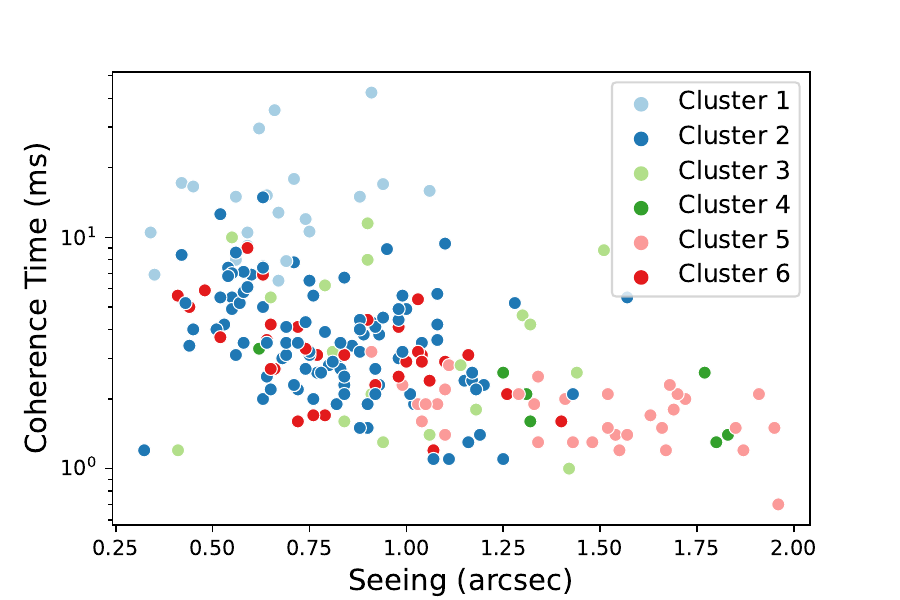}  
            \caption{}
            \label{clustering_coherence_seeing}
        \end{subfigure}
        \begin{subfigure}[b]{.46\textwidth}
            \centering
            \includegraphics[width=\textwidth]{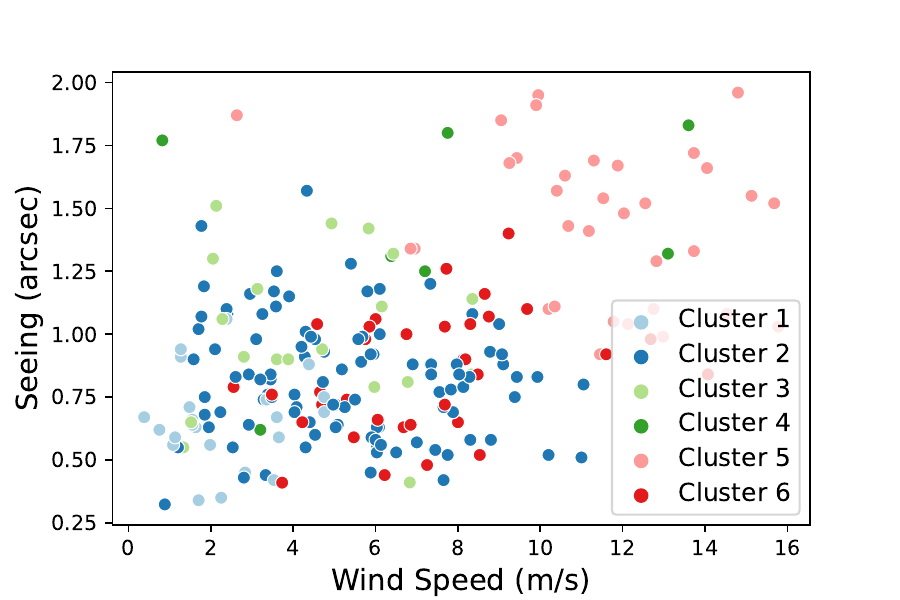}  
            \caption{}
            \label{clustering_seeing_wind}
        \end{subfigure}

        \begin{subfigure}[b]{.46\textwidth}
            \centering
            \includegraphics[width=\textwidth]{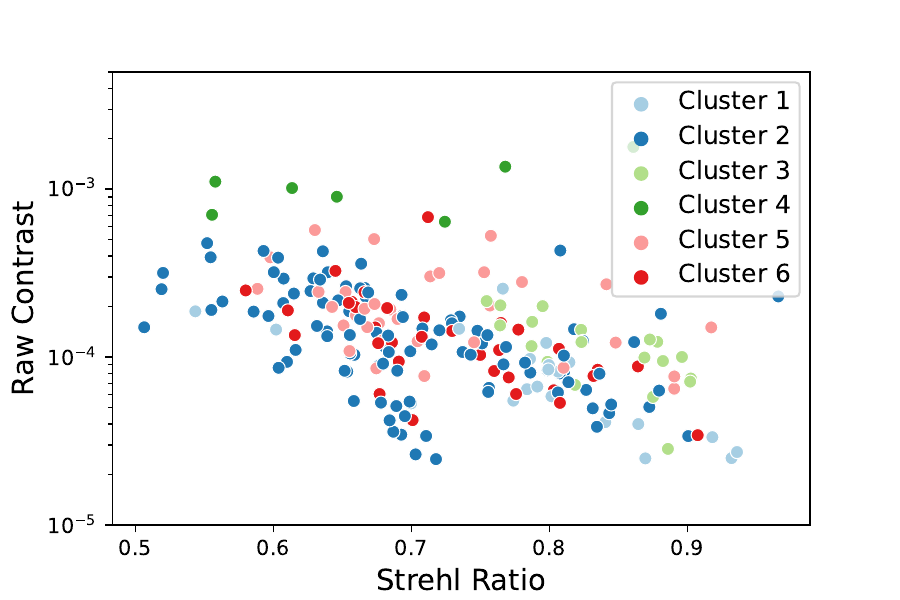}  
            \caption{}
            \label{clustering_strehl_contrast}
        \end{subfigure}
        \begin{subfigure}[b]{.46\textwidth}
            \centering
            \includegraphics[width=\textwidth]{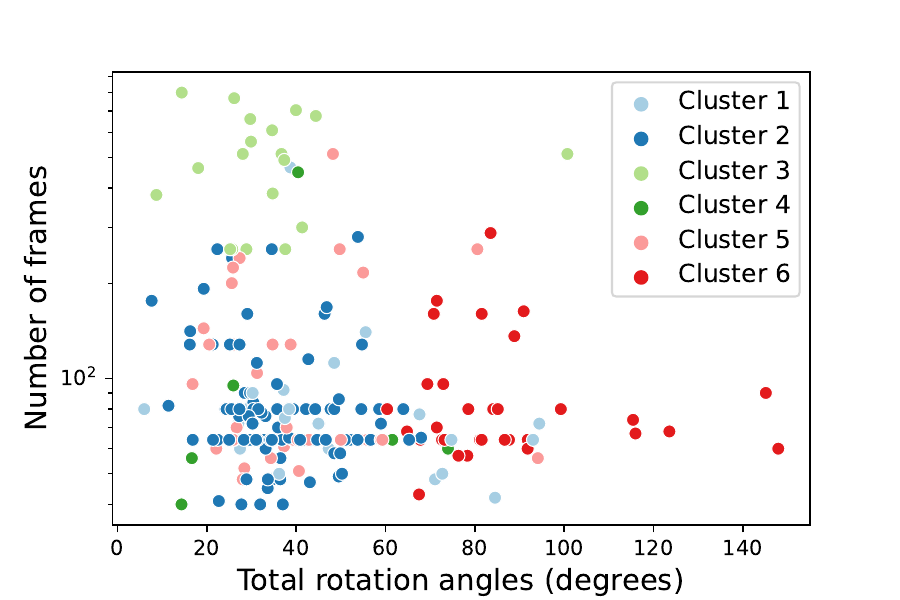}  
            \caption{}
            \label{clustering_frames_rotation}
        \end{subfigure}

        \caption{Projection of cluster distributions across parameters: seeing, coherence time, wind speed, raw contrast, Strehl ratio, total parallactic angles, and number of frames.}
        \label{figure_clusters}
    \end{center}
    
\end{figure*}

To visualize the clustering effectiveness based on selected parameters, we applied Principal Component Analysis (PCA) for dimensionality reduction (see Appendix~\ref{appendix_clustering}), enabling clear separation of clusters in the latent space. For a more physically grounded interpretation of the nature of the various clusters, Fig.~\ref{figure_clusters} presents their visualization based on key environmental parameters, revealing distinct traits for each cluster:
\begin{itemize} 
    \item Cluster 1 (Fig.~\ref{median_frames}a, 11\% of observations) is associated with low wind speed (Fig.~\ref{clustering_seeing_wind}) and high coherence time (Fig.~\ref{clustering_coherence_seeing}), suggesting the presence of the low-wind effect;
    \item Cluster 2 (Fig.~\ref{median_frames}b, 48\% of observations) represents observations under generally fair to good conditions;
    \item Cluster 3 (Fig.~\ref{median_frames}c, 8\% of observations) is characterized by a large number of frames (Fig.~\ref{clustering_frames_rotation});
    \item Cluster 4 (Fig.~\ref{median_frames}d, 3\% of observations) is defined by low raw contrast, low coherence time, and high seeing (Figs. \ref{clustering_coherence_seeing}, \ref{clustering_strehl_contrast}); 
    \item Cluster 5 (Fig.~\ref{median_frames}e, 15\% of observations) shows high wind speed, high seeing, and low coherence time (Figs.~\ref{clustering_coherence_seeing}, \ref{clustering_seeing_wind}), which makes it susceptible to strong wind driven halo; 
    \item Cluster 6 (Fig.~\ref{median_frames}f, 15\% of observations) is associated with large parallactic angle ranges (Fig.~\ref{clustering_frames_rotation}).
\end{itemize}
This successful clustering of the data into these meaningful categories allowed us to analyze noise distributions and detection limits based on well-defined observational conditions, and enabled a more targeted approach to interpreting our results.

\section{The RSM algorithm} \label{section_rsm_algorithm}

The RSM is a well-established econometric algorithm that was successfully adapted to HCI by \citet{2020A&A...633A..95D} to enhance the detection of faint companions at small angular separations from their host stars. In the RSM algorithm, patches of pixels from derotated, residual frames after PSF subtraction are used to construct time series that capture spatial and temporal noise behavior at a given separation. When a planetary signal is present, these time series exhibit significant deviations from the noise patterns across the pixel patch and instead follow a planetary model that is both spatially and temporally structured. The algorithm models the data with two regimes: a noise regime, describing the mean and variance of noise at a given separation, and a planetary regime, incorporating the noise profile to which is added a model of the PSF whose intensity, $\beta$, is 
tuned. Each central pixel of the patch in a given annulus is assigned a likelihood of belonging to either regime. The RSM algorithm employs a Markov Chain model, where the probability of a pixel belonging to the planetary regime depends on the likelihood of the pixel-centered patch alignment with a planetary model, the probability of adjacent pixels, and a transition probability that connects these points. This integration of spatial and temporal constraints, combined with the transition probabilities, enables the algorithm to reliably distinguish speckles from true astrophysical signals. Additionally, the RSM algorithm utilizes outputs from various PSF subtraction techniques to construct a unified time series for each annulus. By drawing on the temporal noise diversity across different post-processing methods, this approach enhances the sensitivity to faint companions. 

RSM has demonstrated significant advancements over current post-processing techniques, as highlighted in multiple studies \citep{2020A&A...633A..95D,2021A&A...646A..49D, 2022A&A...666A..33D}. Furthermore, RSM shows excellent performance compared to other post-processing algorithms, achieving the lowest false positive rate in the Exoplanet Imaging Data Challenge \citep[EIDC;][]{10.1117/12.2574803}, underscoring its reliability in distinguishing planetary signals from noise.

One of the most sensitive aspects of the RSM algorithm involves selecting optimal parameters for both PSF subtraction techniques and RSM itself,  including the planetary flux multiplicative factor $\delta$ \citep[where planetary flux is expressed as a multiple of the standard deviation of pixel intensity at a given separation, $\beta = \delta \sigma$, see equation 1 in][]{2020A&A...633A..95D}, PSF model crop size, and the region used for noise estimation within an annulus. \citet{2021A&A...656A..54D} introduced an automated optimization process within RSM to identify these optimal settings by maximizing contrast for PSF subtraction parameters and minimizing false positives for RSM parameters using the reversed parallactic angles, thereby enhancing the algorithm’s reliability in exoplanet detection.  In this study, we employ RSM with a simplified parameter set, as recommended by \citet{2022A&A...666A..33D}. In the following, we refer to the value of each pixel in the final RSM map -- previously defined as the RSM probability -- as the RSM score, to avoid any potential ambiguity.

Our study incorporates several PSF subtraction techniques: annular principal component analysis \citep[APCA;][]{2013A&A...559L..12A}, locally optimized combination of images \citep[LOCI;][]{2007ApJ...660..770L}, non-negative matrix factorization \citep[NMF;][]{Ren_2018ApJ...852..104R}, as well a forward-model version of KLIP \citep[FM-KLIP;][]{2016ApJ...824..117P} and LOCI \citep[FM-LOCI, see more in ][]{2021A&A...646A..49D}. RSM scores are computed over a radial range of 0\farcs11  (9 pixels) to 1\farcs1  (90 pixels), except for forward-model techniques, which are restricted to a region around 0\farcs35 (30 pixels) due to their computational cost and their tendency to produce results similar to APCA at larger separations. RSM scores are estimated using the forward-backward approach \citep[as defined in][]{2021A&A...646A..49D}, which integrates information from both past and future frames in the ADI time series. Parameters are optimized via Bayesian optimization, as described in \citet[][]{2021A&A...656A..54D}.

To reduce computational load and mitigate potential issues with the RSM parameter optimization on certain pathological datasets, we propose to determine the optimal parameters for the cluster centers identified in Sect.~\ref{section_environmental_conditions} and to apply these settings consistently across all datasets within each cluster. Table~\ref{tab:RSMparams} in Appendix~\ref{table_optimal_parameters_RSM} details the specific parameters for each cluster. To assess the reliability of this approach, Appendix~\ref{appendix_similarities_RSM} compares the cluster center parameters with those computed individually for each dataset, providing insights into their alignment and the robustness of the clustering methodology. We exclude from the optimization process the selection of the optimal combination of PSF subtraction techniques within the auto-RSM framework, as delivered by the opti-combination method described in \citet{2021A&A...656A..54D}. Instead, we keep this parameter adjustable and explore how varying combinations of PSF subtraction techniques impact noise distributions and contrast curves across separations, as discussed in Sect.~\ref{section_contrast_curves}.

\section{Noise behavior in RSM} \label{section_noise_behavior_in_RSM}

In HCI, detection limits are commonly set at a $5 \sigma$ level, corresponding to a false alarm probability (FAP) of $3 \times 10^{-7}$ based on a Gaussian noise assumption in processed frames \citep{2014ApJ...792...97M}. However, this approach may not accurately reflect the true noise characteristics in HCI datasets. Alternative approaches based on, for example, Laplacian distributions \citep{2019MNRAS.487.2262P, 2020A&A...633A..95D} have been proposed to address this limitation, especially at small separations. Moreover, methods such as those in \citet{2018AJ....155...19J}, \citet{2023AJ....166...71B}, and \citet{2024A&A...692A.126D} avoid assuming any specific noise model, instead using adaptive techniques to derive thresholds directly from data, reducing the reliance on potentially mismatched distributions.

In the context of RSM, fitting a theoretical distribution to the noise in the final RSM maps (hereafter referred to as RSM noise) is challenging due to the nonlinear dependency of the final RSM score on the likelihood functions and transition probabilities, leading to the distortion of any pre-assumed distribution. \citet{2021A&A...646A..49D} defines the detection threshold for the contrast curve as the brightest false positive (FP) in the final detection map. As noted in Appendix~\ref{appendix_completeness}, this threshold is broadly consistent with those defined in this section. However, it lacks robustness in the presence of astrophysical signals and is not designed to detect them. An alternative approach explored by \citet{2022A&A...666A..33D} , based on reversed parallactic angles, was also considered as a potential threshold applicable to both detection maps and contrast curves. However, it posed challenges by revealing a high number of false positives, ultimately limiting its utility as a reliable detection criterion.

In this study, we investigate a new, consistent threshold applicable both for generating contrast curves and identifying candidate companions in RSM. Here, we examine two methods for defining a detection threshold, either (i) by fitting an empirical distribution to independent RSM noise realizations in the final RSM map (Sect.~\ref{subsection_lognormal_distribution}), or (ii) by balancing false positives and true positives while maximizing the F1 score, without assuming a specific RSM noise distribution (Sect.~\ref{subsection_F1_score}). We also explore how varying observing conditions influence the RSM noise distribution and the resulting threshold. Finally, we evaluate the strengths and limitations of each threshold within the context of RSM.

\subsection{Fitting a lognormal distribution to the RSM noise} \label{subsection_lognormal_distribution}

In this method, we take advantage of the large amount of data in the SHINE F150 survey to build distribution samples at each angular separation. Following \citet{2023AJ....166...71B}, we treat independent pixel realizations, i.e., the central pixels of non-overlapping resolution elements, as distribution samples. Noise analysis at a given separation across the entire survey shows multiple RSM noise distributions, leading us to analyze realizations per cluster and per combination of PSF subtraction techniques in RSM, at each considered angular separation. 

The final RSM maps contain only positive values, leading to a noise distribution that is not centered on zero and exhibits positive skewness.
After testing multiple distributions as shown in Appendix~\ref{appendix_lognormal_distribution}, we selected the lognormal distribution for its minimal parameter requirements and its consistency with the shape of the RSM noise distribution across separations and clusters. To define a detection threshold, we set a $3 \times 10^{-7}$ FAP for the lognormal distribution, aligning with the conventional $5\sigma$ threshold in Gaussian statistics. To prevent bias when fitting the distribution, we exclude scores above 0.1 from the distribution samples, as these correspond to bright signals. 

\begin{figure}[t]
    \centering
    \begin{subfigure}[b]{.23\textwidth}
        \centering        \includegraphics[width=\textwidth]{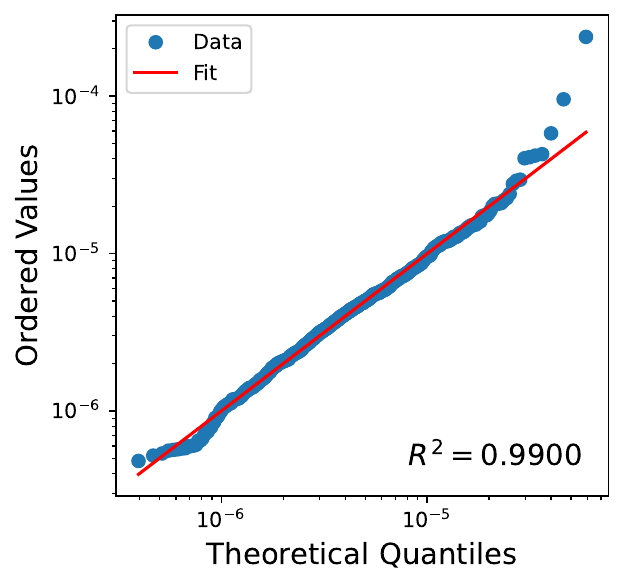}  
        
        \label{QQ_plot_5}
    \end{subfigure}
    \begin{subfigure}[b]{.23\textwidth}
        \centering
        \includegraphics[width=\textwidth]{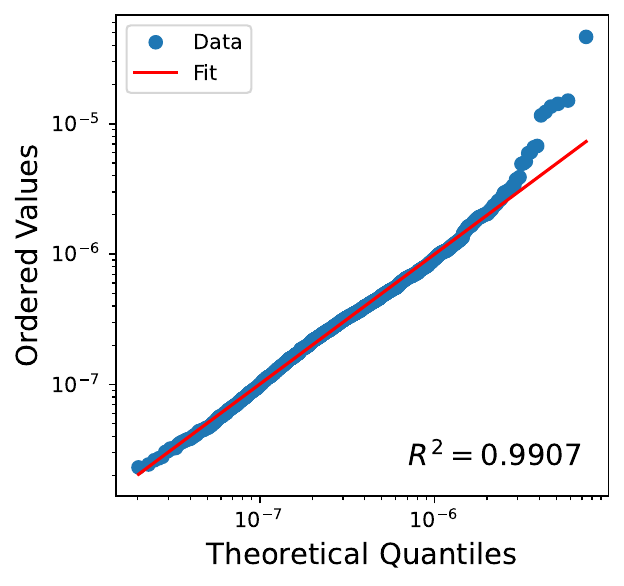}  
        
        \label{QQ_plot_15}
    \end{subfigure}
    \caption{Quantile-quantile plot of the RSM noise in Cluster 1 at $0\farcs25$ (left) and $1\arcsec$ (right) using APCA. The plot compares the RSM noise distribution to the lognormal distribution, excluding values above the $3 \times 10^{-7}$ FAP of the fit. The coefficient of determination ($R^2$) is also displayed, indicating a good fit.}
    \label{Q-Q plots}
\end{figure}

Figure~\ref{Q-Q plots} presents quantile-quantile (Q-Q) plots for cluster 1 at small (left) and large (right) separations using RSM with APCA. We compare the logarithm of independent sample values from all observations in cluster 1 at 
0\farcs25 and 1\arcsec\ with the corresponding fitted distribution values, considering only those below the $3 \times 10^{-7}$ threshold. 
The Q-Q plots exhibit an approximately linear trend in both cases, indicating a strong fit, even though the extremities of the distribution show some outliers as it is often the case in Q-Q plots. Moreover, the coefficient of determination ($R^2$), representing the Pearson correlation between quantiles, is close to 1, further confirming the goodness of fit.

\begin{figure}[t]
    \centering
    \begin{subfigure}[b]{.5\textwidth}
        \centering
        \includegraphics[width=\textwidth]{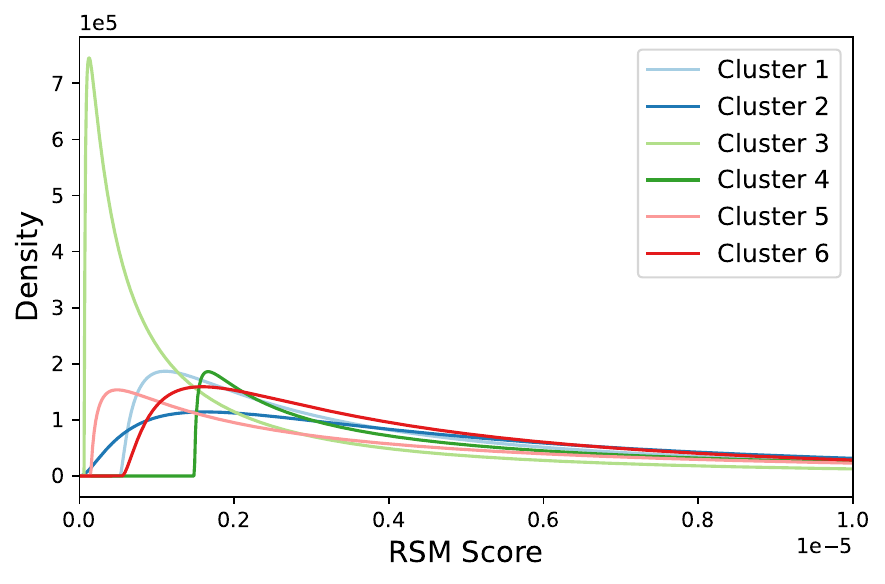}  
        \caption{}
        \label{diff_distributions}
    \end{subfigure}
    \begin{subfigure}[b]{.5\textwidth}
        \centering
        \includegraphics[width=\textwidth]{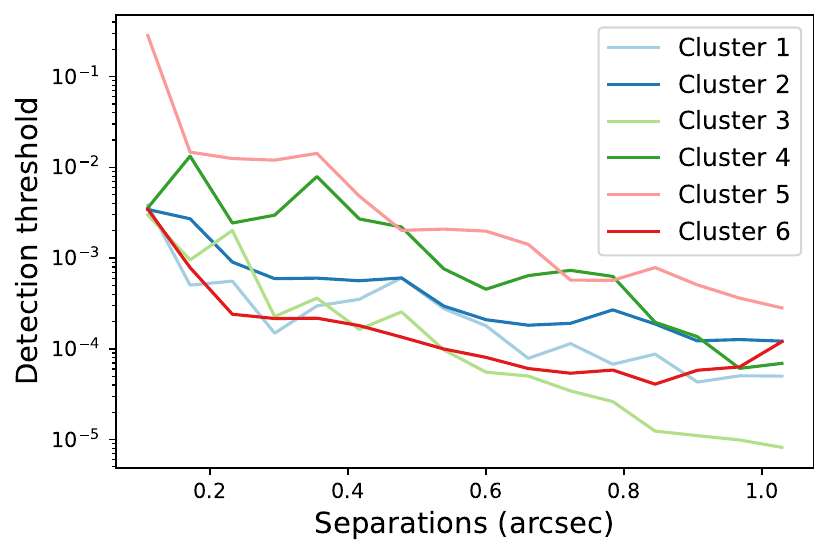}
        
        \caption{}
        \label{diff_thresholds}
    \end{subfigure}
    \caption{\textit{Top}. Lognormal distributions obtained by fitting RSM noise histograms at $0\farcs25$ for various observing conditions (clusters) using RSM with APCA. \textit{Bottom}. Detection thresholds derived from the lognormal distributions at $3\times 10^{-7}$ FAP, as a function of angular separation and of observing conditions.}
    \label{lognormal_different_clusters}
\end{figure}

In Figure \ref{lognormal_different_clusters}, we show how varying observing conditions impact the RSM noise distribution and, consequently, the chosen detection threshold. Figure \ref{diff_distributions} presents the lognormal fits for different clusters at $0\farcs25$ using RSM with APCA. Cluster 3, featuring a large number of frames per target, shows a narrower distribution and noise tail, while cluster 4, with mostly poor-quality observations, shows a broader tail and wider distribution, indicating a higher detection threshold. This plot highlights the varying RSM noise behaviors across observations under different conditions. Figure \ref{diff_thresholds} illustrates the variations of the $3\times 10^{-7}$ FAP threshold across clusters and separations, highlighting the advantage of setting separate thresholds per separation to accommodate distinct noise properties. It also emphasizes the benefit of adjusting thresholds for observations under varying conditions. This approach, consistent with predictions, results in lower thresholds for Cluster 3 and higher ones for Cluster 4 and 5. This analysis underscores the importance of considering observing conditions when setting thresholds, as applying a single threshold across all datasets could lead to significant underestimations or overestimations of the detection thresholds, as evidenced by the almost two orders of magnitude difference in thresholds between the best and worst clusters.

\subsection{Threshold using maximum F1 score}
\label{subsection_F1_score}

In this section, we explore the possibility of defining a detection threshold without assuming a specific RSM noise distribution, an approach particularly relevant for samples containing a limited number of observations, where robust noise histograms cannot be built, especially for small angular separations. Following \citet{2024A&A...692A.126D}, we define detection thresholds by empirically maximizing the F1 score at each angular separation through fake companion injections. The F1 score is defined as follows:
\begin{equation}
 \text{F1 score} = \frac{2\mathrm{TP}}{2\mathrm{TP} + \mathrm{FP} + \mathrm{FN}} \: ,
\end{equation}
with TP the true positives, FP the false positives, and FN the false negatives. Maximizing the F1 score ensures an optimal balance between true positive rate and false positive rate.

\begin{figure}[t]
    \centering
    \includegraphics[width=0.45\textwidth]{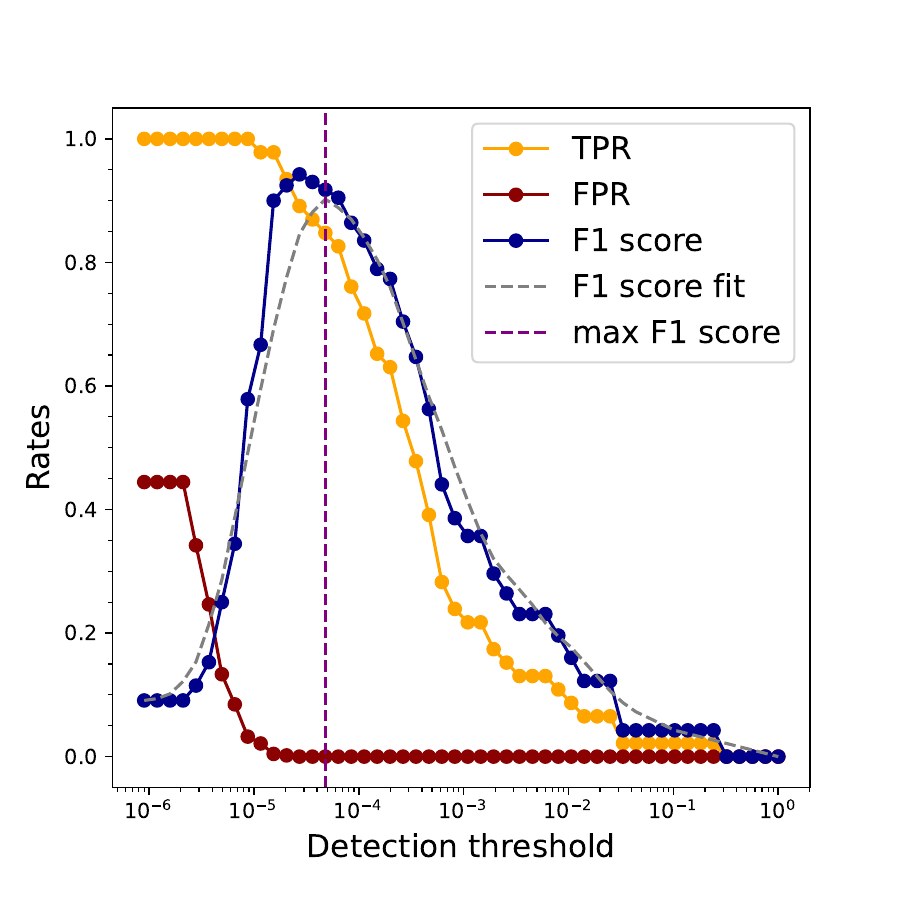}
    \caption{Variation of the TPR, FPR, and F1 score for the cluster 1 center at $0\farcs7$. The dashed purple line marks the threshold that maximizes the fitted F1 score curve (dark blue curve), balancing high true positive (orange curve) recovery with minimal false positives (dark red curve).}
    
    \label{F1 score}
\end{figure}

To compute the false positives, all independent pixel realizations at each separation are considered as devoid of circumstellar signal (pure noise). Varying the threshold, we calculate the number of aperture values exceeding the threshold, which represent the number of false positives in the considered annulus. The false positive rate (FPR) is then defined as the ratio of these false positives to the total number of independent realizations.
For true positives, we follow \citet{2014ApJ...792...97M} by first estimating noise levels in the residual cube produced by APCA. We then inject synthetic point sources with fluxes between two and three $\sigma$ of the measured noise. An injected source is considered recovered if its RSM score surpasses the considered detection threshold, and the number of such recoveries within the annulus defines the true positives. The optimal threshold is determined by maximizing the F1 score, ensuring a balance between high true positive recovery and minimal false positives.

Figure \ref{F1 score} shows the evolution of TPR, FPR, and the F1 score for varying thresholds, applied to the cluster 1 center (corresponding to the BD-15~705 dataset) using RSM with APCA at $0\farcs7$ separation. The figure highlights the fact that the optimal F1 score favors minimal FPR and high TPR, thereby effectively distinguishing noise from potential true positives.
This metric is highly dataset-specific as it relies heavily on the false positive rate and the true positive rate, the latter also depending on the flux of injected signals, which is influenced by the noise distribution. Consequently, this approach captures the unique characteristics of each dataset and aligns the threshold closely with the specific noise properties of the data. This technique, however, has several limitations. First, while it minimizes the false positive rate, it does not explicitly control the acceptable number of false positives per annulus. Additionally, it assumes all independent realizations are false positives, which may not hold if a real signal is present. In such cases, the maximum F1 score metric tends to exclude the true signal by minimizing false positives. One solution is to remove bright signals by injecting negative fake companions to avoid bias. Finally, this approach requires noise computation, fake companion injections, and a complete RSM processing for each separation, making it computationally expensive.

\subsection{Comparison and reliability of the detection thresholds} \label{reliability threshold}

\begin{figure}[t]
    \centering
    \includegraphics[width=0.55\textwidth]{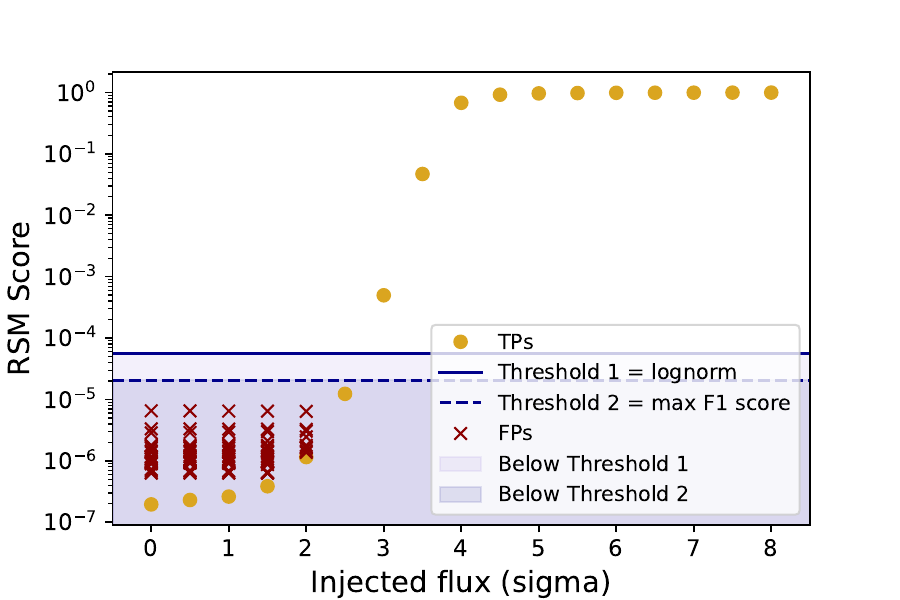}
    \caption{Comparison of detection thresholds with false positives and injected companions for the cluster 1 center at an angular separation of $0\farcs6$ and showing the RSM scores of injected fake companions (gold dots), false positives (red crosses), the maximum F1 score threshold (dashed blue line), and the lognormal threshold (solid blue line).}
    \label{injections_for_thresholds}
\end{figure}

\begin{figure}[t]
    \centering
    \includegraphics[width=\linewidth]{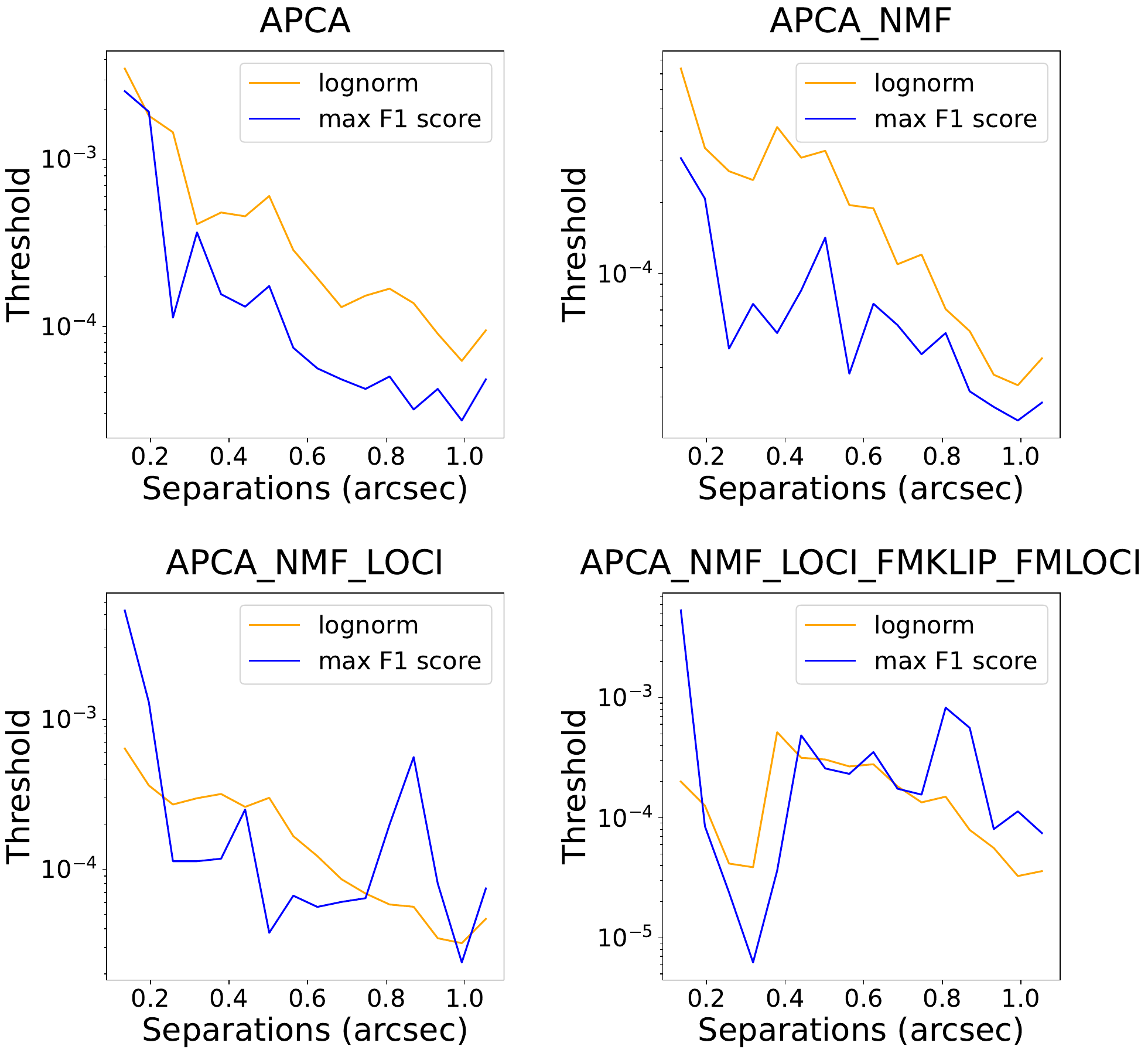}
    \caption{Median detection thresholds for the six different clusters using the $3 \times 10^{-7}$ FAP under lognormal distribution (orange), compared with the threshold derived from the maximum F1 score on the cluster centers (blue), for different combinations of PSF subtraction techniques as described in the plot titles.}
    \label{comparison_thresholds}
\end{figure}

To evaluate the reliability of detection thresholds, we examine their proximity to false positives for the cluster center of cluster 1. Figure~\ref{injections_for_thresholds} compares the RSM scores of fake companions injected at an angular separation of $0\farcs6$, across flux levels ranging from 0 to 8$\sigma$, with the false positives detected at that separation and the two associated detection thresholds.
For each injection, we computed the number of independent apertures within the considered annulus whose mean values exceeded the recovered flux, treating them as false positives. Both the maximum F1 score threshold and the lognormal threshold lie above the false positives, with the F1-score threshold closely aligning with the $2.5\sigma$ injected fake companion, near the noise-dominated region. The lognormal threshold, slightly higher at approximately $3\sigma$, indicates a more conservative detection criterion.
Figure~\ref{comparison_thresholds} shows the median detection thresholds across all clusters, with each lognormal threshold computed individually for each cluster considering all its members, and the maximum F1 score thresholds computed for the cluster center datasets. The dip in thresholds below $0\farcs4$ in the bottom-right panel of Fig.~\ref{comparison_thresholds}  corresponds to the influence of forward modeling techniques (FM KLIP and FM LOCI), which are active within this separation range. This comparison highlights the differences between the two methods, with the lognormal threshold being generally more conservative. The larger variability in the detection threshold derived from the maximum F1 score across separations reflects its sensitivity to outliers in different annuli.

While it arguably provides the most relevant thresholds for a given dataset, the maximum F1 score is computationally demanding to apply on each dataset of the survey, and too specific to generalize across an entire cluster (see more about that in \mbox{Appendix}~\ref{appendix_F1_centers}). Conversely, the detection thresholds derived from a lognormal distribution are more conservative, computationally efficient, and reflect the noise behavior under similar observing conditions. However, they rely on a predefined distribution, large sample sizes, and do not capture dataset-specific nuances.
For the present study, focusing on a large survey, we adopt the threshold derived from the $3 \times 10^{-7}$ FAP of the lognormal distribution, and derive it at each separation, for each combination of PSF subtraction technique, and for each cluster. We recommend using the F1 score threshold for smaller samples, while noting its limitations described in Sect.~\ref{subsection_F1_score}.

\section{Sensitivity limits} \label{section_contrast_curves}

In this study, we adopt the 50\% completeness curve described in \citet{2021A&A...646A..49D}, also known as contrast curves, to assess the sensitivity limits to point-like sources. Such curves indicate the flux at which 50\% of injected signals are detected above the selected threshold, here the $3 \times 10^{-7}$ FAP from the lognormal distribution, which corresponds to the $5 \sigma$ contrast curve when assuming Gaussian noise. Here, we analyze the effects of threshold selection and PSF subtraction technique combinations on contrast curves, define the optimal RSM contrast curve, and evaluate its reliability.

\subsection{Influence of threshold and PSF subtraction techniques} \label{subsection_contrast_curves}

\begin{figure}[t]
    \centering
    \includegraphics[width=\linewidth]{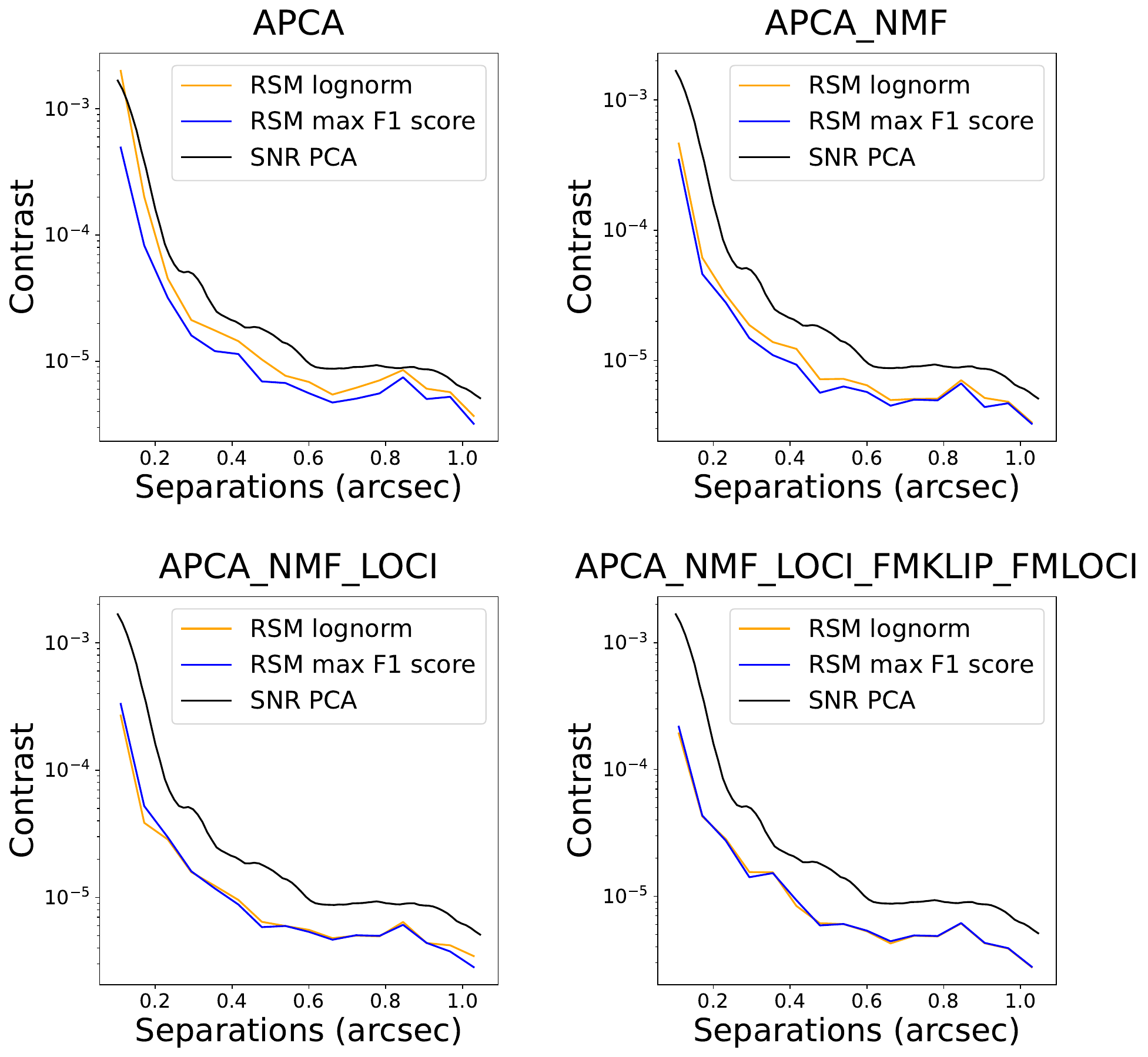}
    \caption{Contrast comparison at 50\% completeness using thresholds based on the maximum F1 score (blue), the lognormal RSM noise assumption (orange), and the $5\sigma$ APCA contrast curve (black) across different PSF subtraction techniques applied to the center of cluster 1.}
    \label{contrasts_comparisons_thresholds}
\end{figure}
To evaluate how the chosen threshold impacts contrast curves, Fig.~\ref{contrasts_comparisons_thresholds} presents a comparison for the center of cluster 1. It includes contrast curves derived using the maximum F1 score threshold (blue) and the lognormal threshold (orange) across four combinations of PSF subtraction techniques: APCA, APCA-NMF, APCA-NMF-LOCI, and APCA-NMF-LOCI-FMKLIP-FMLOCI. A standard $5\sigma$ contrast curve based on full-frame PCA with five principal components is shown in black for comparison. The RSM contrast curves show a good level of consistency between the two ways to define the detection threshold. This is especially the case when multi-technique combinations are used (see bottom panels in Fig.~\ref{contrasts_comparisons_thresholds}), while the lognormal threshold leads to slightly worse sensitivity than the F1-score one when RSM is performed with APCA only (Fig.~\ref{contrasts_comparisons_thresholds} top left) due to the higher threshold in that case. Consistency was also found between the lognormal contrast curves derived here and the contrast curves based on the appearance of a first false positive proposed in \citet{2022A&A...666A..33D} for the SHARDDS survey (see Appendix~\ref{appendix_completeness}).

\begin{figure*}[t]
    \centering
    \includegraphics[width=\linewidth]{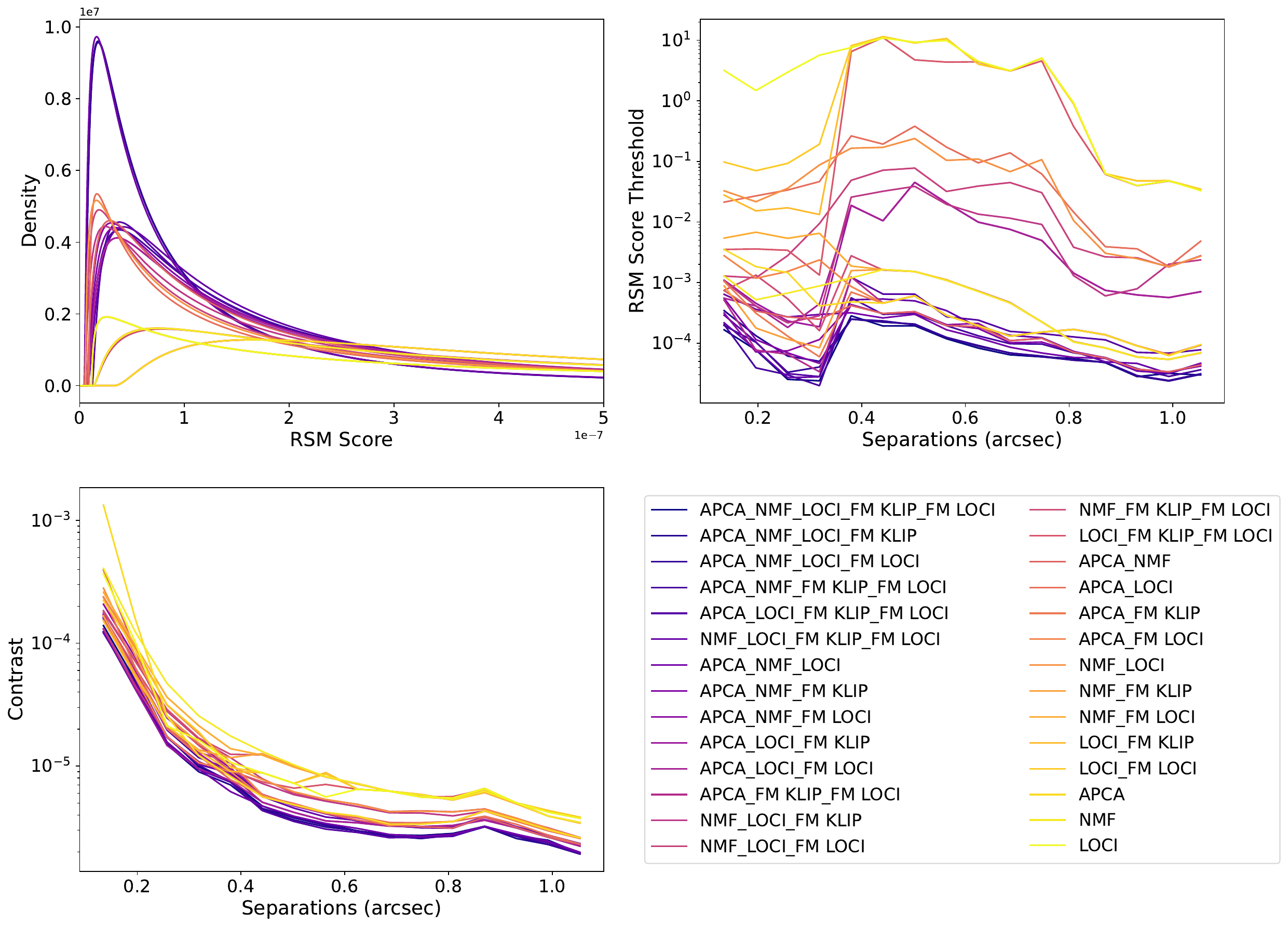}
    \caption{Effect of different combinations of PSF subtraction techniques in RSM on the fitted lognormal distributions, detection thresholds, and contrast curves. \textit{Top left:} Difference of lognormal fit across different PSF subtraction techniques with colors ranging from dark violet for multi combinations to yellow for singular combinations. \textit{Top right:} Distribution of the $3 \times 10^{-7}$ FAP under lognormal distribution for the different combinations across all separations. 
    \textit{Bottom left:} Contrast curve delivered by the different combinations with 50\% completeness.}
    \label{combinations}
\end{figure*}

The choice of which PSF subtraction techniques to combine is an important aspect of the RSM algorithm, which is not straightforward to optimize. Instead of relying on a single combination identified as optimal by the RSM framework, we propose to systematically evaluate the performance of various combinations of PSF subtraction techniques. We analyze the differences in noise distributions in the final frames, the corresponding thresholds, and the contrast curves achievable with each combination. Figure~\ref{combinations} examines these effects across the 28 possible combinations of PSF subtraction techniques, ranging from single PSF subtraction techniques (in yellow) to multi-combinations (in dark violet). Figure \ref{combinations} (top left) displays the median lognormal fits for each combination, clearly showing narrower noise distributions for multi-combinations and broader distributions for single techniques, indicating improved performance with combined techniques. Figure \ref{combinations} (top right) presents the thresholds derived from these fits across all combinations and separations. Noise thresholds, on average, range from approximately $10^{-4}$ for multi-combinations to nearly ten for LOCI -- exceeding RSM’s maximum score. This renders LOCI non-physical and unreliable when used alone, a consequence of its high false positive rate in RSM for many datasets. Figure \ref{combinations} (bottom left) illustrates the contrast curves for the various combinations, showing an improvement by a factor of three for multi-combinations. Contrast values with lognormal thresholds exceeding 0.1 in RSM were excluded from the bottom-left panel due to difficulties in achieving convergence. This further demonstrates that incorporating multiple PSF subtraction techniques in RSM significantly enhances detection limits. Moreover, it highlights RSM’s ability to leverage the diverse noise patterns of different PSF subtraction methods, effectively reducing the misclassification of transient speckles as planetary signals. RSM's efficiency in integrating additional information -- either by combining different PSF subtraction techniques or by increasing the number of frames -- is evident through narrower noise distributions, improved thresholds, and enhanced detection limits (see also Figure~\ref{diff_distributions}).

\subsection{Optimal usage of RSM}

\begin{figure*}[t]
    \centering
    \sidecaption
    \includegraphics[width=12cm]{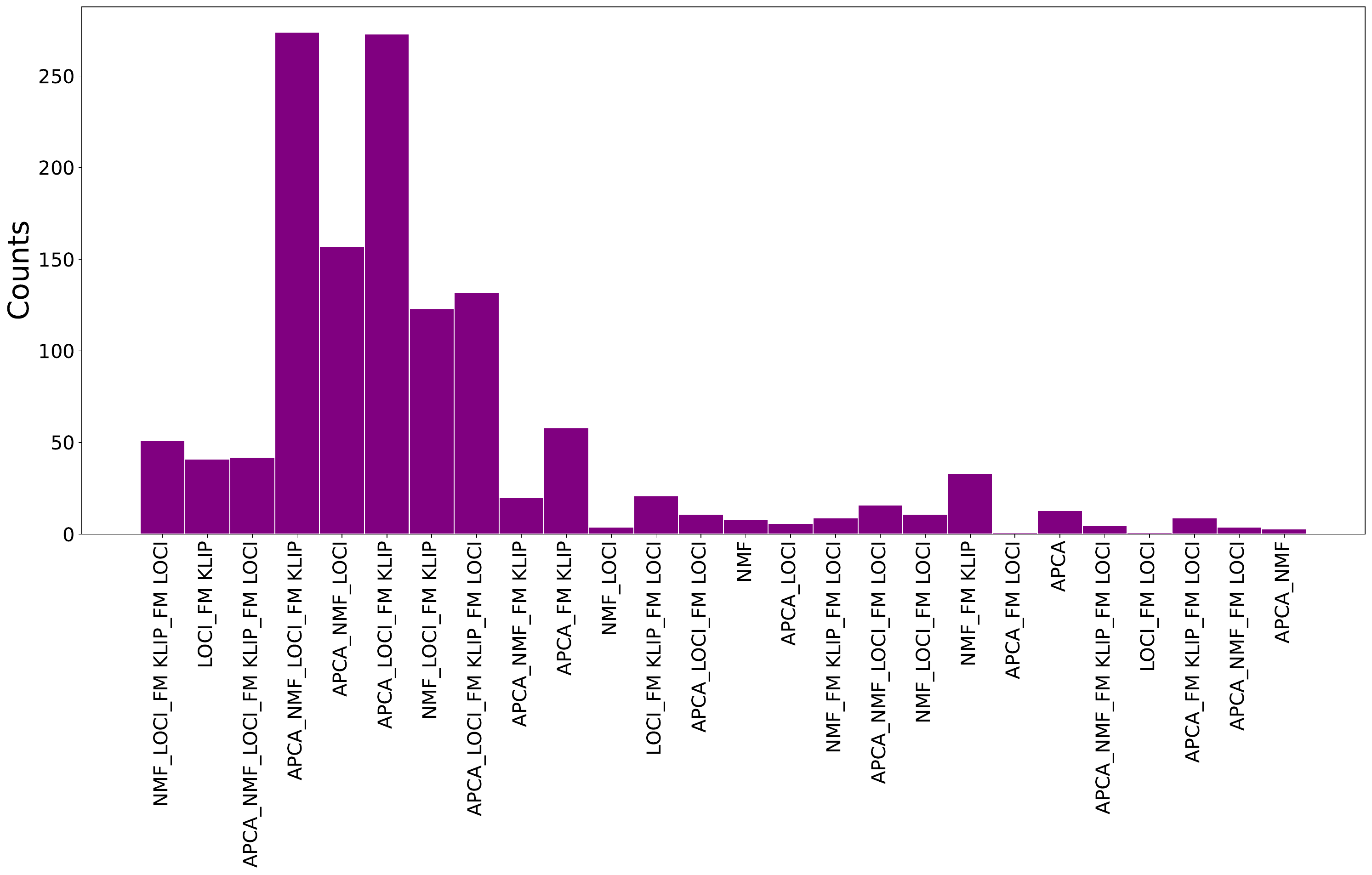}
    \caption{Histogram of RSM combination contributions to the best contrast values across all separations, based on all datasets}
    \label{histogram_best}
\end{figure*}

\begin{figure}[t]
    \centering
    \includegraphics[width=\linewidth]{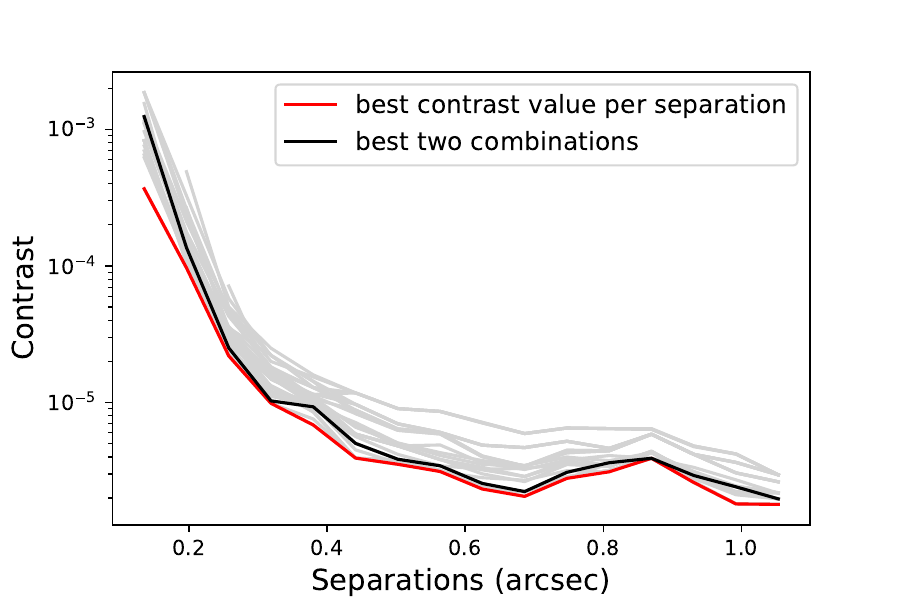}
    \caption{Optimal contrast curve achieved by different combinations of PSF subtraction techniques in RSM.  Gray curves show the median contrast performance for each combination, computed across all datasets in cluster 1. The red curve highlights the best achievable performance at each separation, given by the median of the minimum contrast values across all combinations.  The black curve shows the median of the maximum contrast values delivered by the two best-performing combinations: APCA-NMF-LOCI-FMKLIP and APCA-LOCI-FMKLIP.}
    \label{best_best}
\end{figure}
Although combining multiple techniques yields the best RSM performance, the optimal contrast at each separation does not always result from the same combination.  Figure \ref{histogram_best} presents a histogram illustrating the contributions of various RSM combinations to the best contrast values across all separations, based on all datasets. We identify the best contrast at each separation, shown as the red curve in Fig.~\ref{best_best}. Figure~\ref{histogram_best} indicates that approximately 26 combinations out of the 28 possible combinations of PSF subtraction techniques contribute to the red curve of Fig.~\ref{best_best}, with two combinations standing out due to their significantly higher contributions compared to others: APCA-NMF-LOCI-FMKLIP, and APCA-LOCI-FMKLIP. Since tracking candidates across 26 different RSM maps is complex, we opted to use only the two combinations with the highest contributions. Contrast curves were then computed using the maximum (= worst) contrast value from these two combinations at each separation, and point source candidates were identified if they appeared in both of the combinations above their respective thresholds. This contrast resulting from the aggregation of the best two combinations is represented by the black curve in Fig.~\ref{best_best}. We recommend RSM users adopt the median contrast of these combinations to derive reliable contrast curves and use the corresponding maps to identify candidates. This metric is employed to determine detection limits in Sect.~\ref{section_improvements_in_the_SHINE_survey}.

\begin{figure*}[t]
    \centering
    \begin{subfigure}[b]{.32\textwidth}
        \centering        \includegraphics[width=\textwidth]{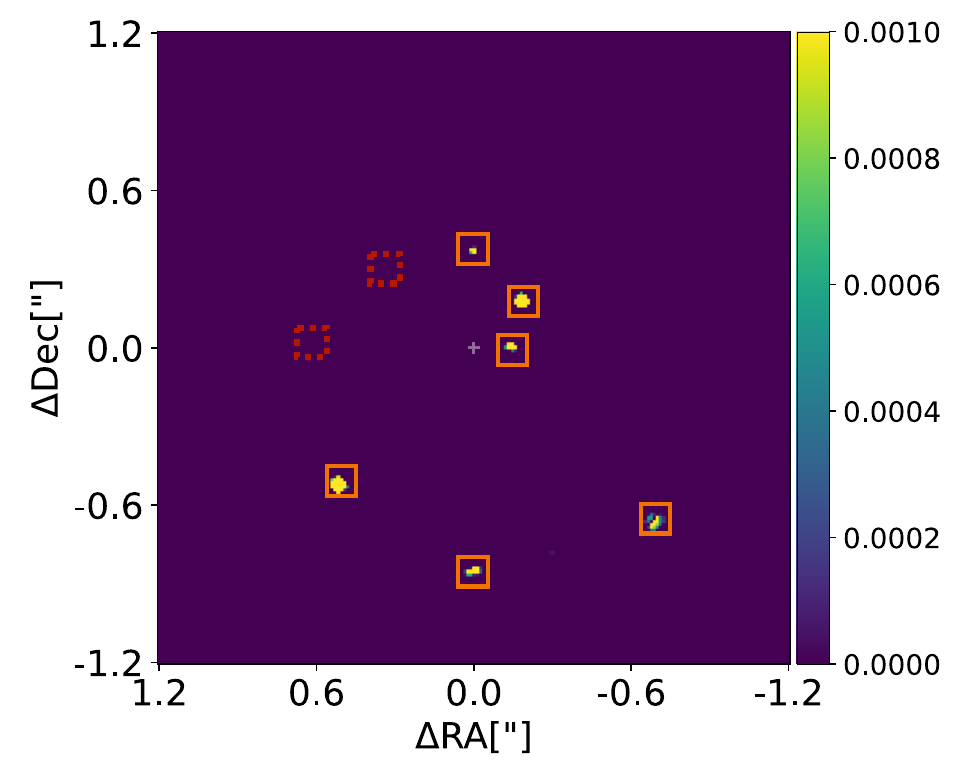}  
        \caption{}
        \label{map_0_1_2_3}
    \end{subfigure}
    \begin{subfigure}[b]{.32\textwidth}
        \centering        \includegraphics[width=\textwidth]{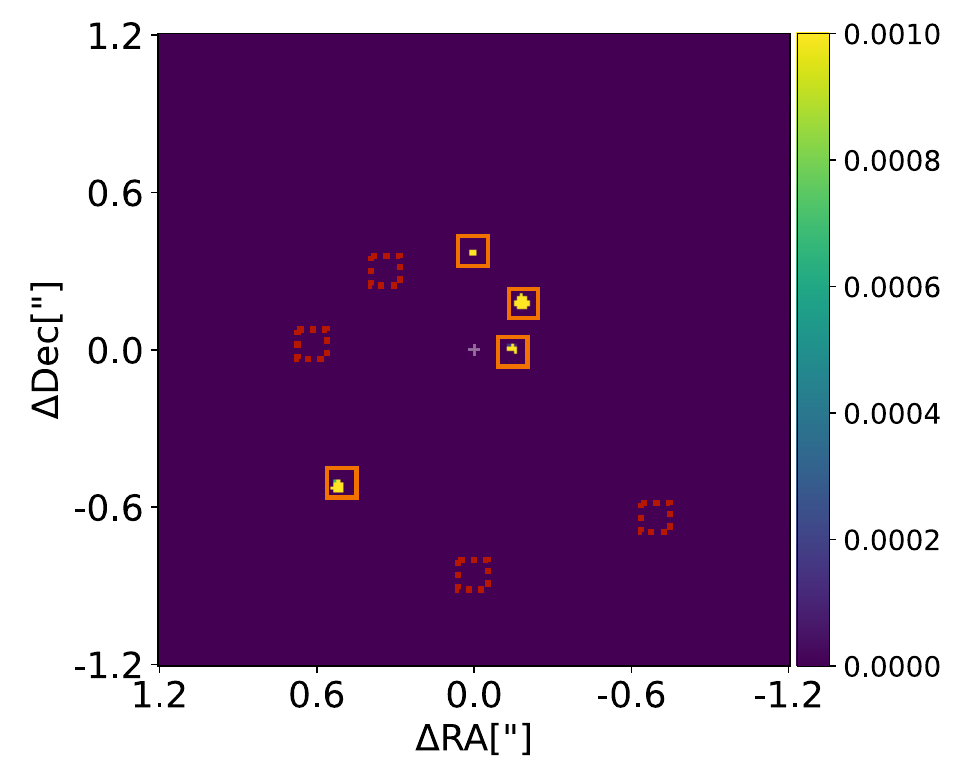}  
        \caption{}
        \label{map_0_2_3}
    \end{subfigure}
    \begin{subfigure}[b]{.30\textwidth}
        \centering        \includegraphics[width=\textwidth]{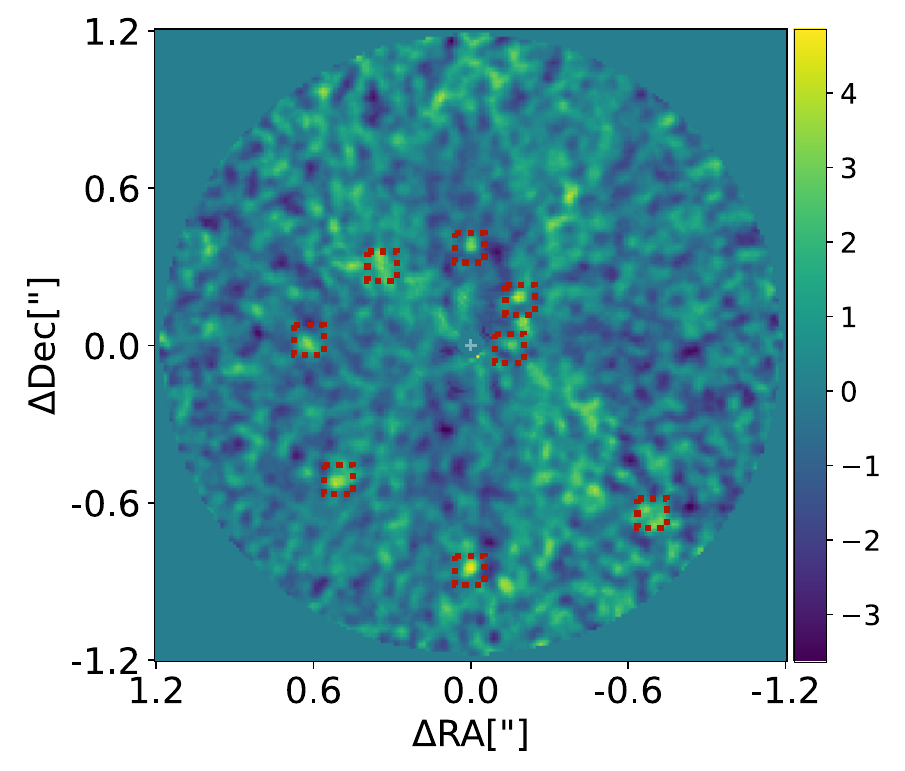}  
        \caption{}
        \label{map_SNR}
    \end{subfigure}
    \caption{Comparison of recovered fake point sources in cluster center 1 using RSM maps with different PSF subtraction technique combinations: (a) APCA-NMF-LOCI-FMKLIP and (b) APCA-LOCI-FMKLIP, compared to the signal-to-noise ratio PCA map with five principal components (c). The RSM maps shown in this figure are thresholded (RSM map minus the separation-dependent threshold). Injected signals recovered above the detection threshold are marked with yellow squares, while unrecovered signals appear in red squares.}
    \label{injection_maps}
\end{figure*}

\begin{figure}[t]
    \centering
    \includegraphics[width=\linewidth]{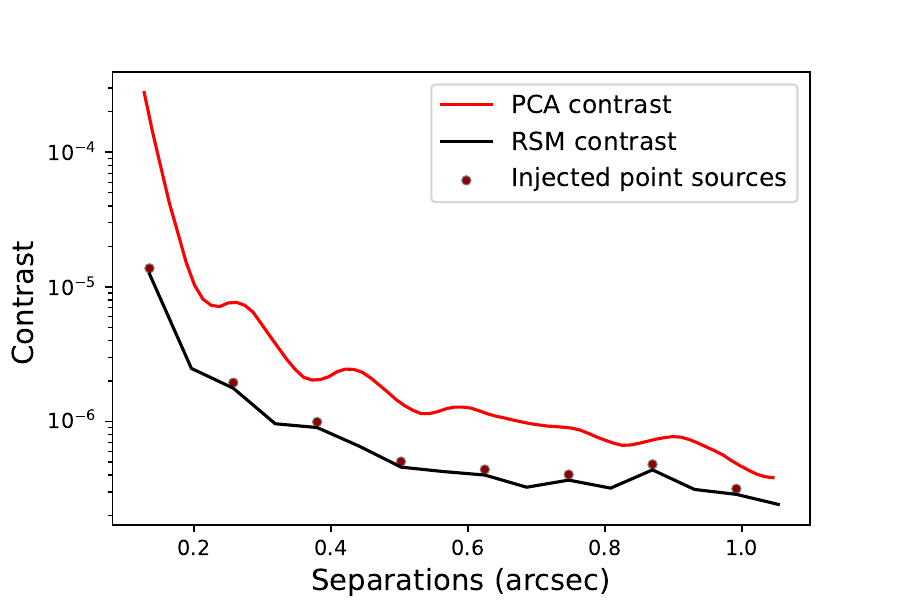}
    \caption{Injected fake point sources compared to the optimal RSM contrast curve (black) and PCA five-component contrast curve (red) for cluster center 1}
    \label{injections}
\end{figure}

\subsection{Reliability of the contrast curves}

Evaluating the adopted contrast curve metric through fake companion injections helps verify the consistency of its core assumptions, including the applied noise threshold, the definition of the median contrast curve, and the expected 50\% recovery rate of injected signals. We inject fake point sources at the predicted sensitivity limit using the VIP package on a cluster center and examine their recovery across different RSM maps. Figure \ref{injection_maps} compares the recovery of injected signals using RSM maps against the signal-to-noise ratio (S/N) map produced by PCA with five principal components, for the center of cluster 1. The flux levels of the injected companions have been chosen to be just above the expected RSM sensitivity limits as illustrated in Fig.~\ref{injections}, which also compares the detection limits of RSM with the full-frame PCA $5\sigma$ contrast curve. The injections are performed at multiple separations, with one injection at each separation and varying position angles. 

As expected, a significant fraction of the injections are successfully detected above the threshold in RSM maps, achieving approximately 50\% recovery consistent with the definition of the contrast curve. Conversely, the same injections all appear well below the $5\sigma$ level in the S/N PCA map. This highlights RSM's ability to deliver deeper detection limits and reliable contrast curves, surpassing conventional post-processing methods in HCI.

\subsection{Improvement to the SHINE survey}
\label{section_improvements_in_the_SHINE_survey}

\begin{figure}[t]
    \centering
    \includegraphics[width=\linewidth]{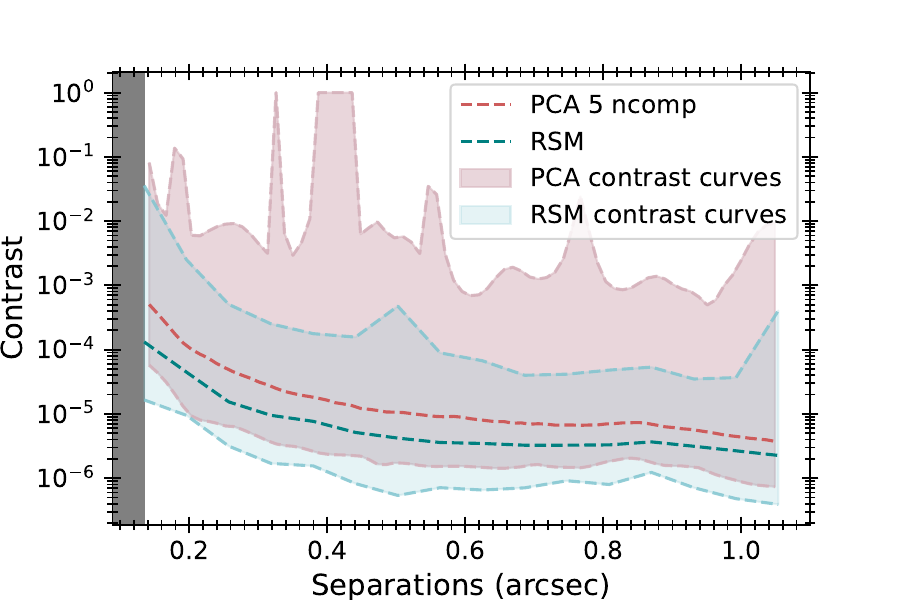}
    \caption{Comparison of RSM and full-frame PCA contrast curves for the F150 SHINE sample. The illustration provides a visualization of all RSM contrast curves (shaded blue) and PCA contrast curves (shaded pink), including their median values, highlighting RSM's significant improvement in detection limits at both small and large separations.}
    \label{Contrast_all}
\end{figure}

To conclude this analysis of sensitivity limits, we present the detection limits for the F150 SHINE sample derived from 213 observations using the detection threshold and modifications outlined in this study. Figure \ref{Contrast_all} shows all RSM contrast curves (shaded blue) against full-frame PCA ADI curves (shaded pink). For consistency within the survey, we do not use a high-pass filter at large separations, although we acknowledge that it could deliver better results in these cases. The PCA-ADI curves may show slight differences from those published in \citet{2021A&A...651A..71L}, reflecting variations in the data reduction process. Nevertheless, our PCA-ADI results align well with those reported in \citet{2023A&A...675A.205C}, as both are based on the same reduction pipeline. The PCA contrast curves display outliers due to bright companions and background star signals.  The median contrast curves indicate RSM's substantial improvement, achieving a factor of two enhancement at 800 mas and a factor of four to five at 135 mas compared to the five-component PCA results.

\begin{figure}[t]
    \centering
    \includegraphics[width=\linewidth]{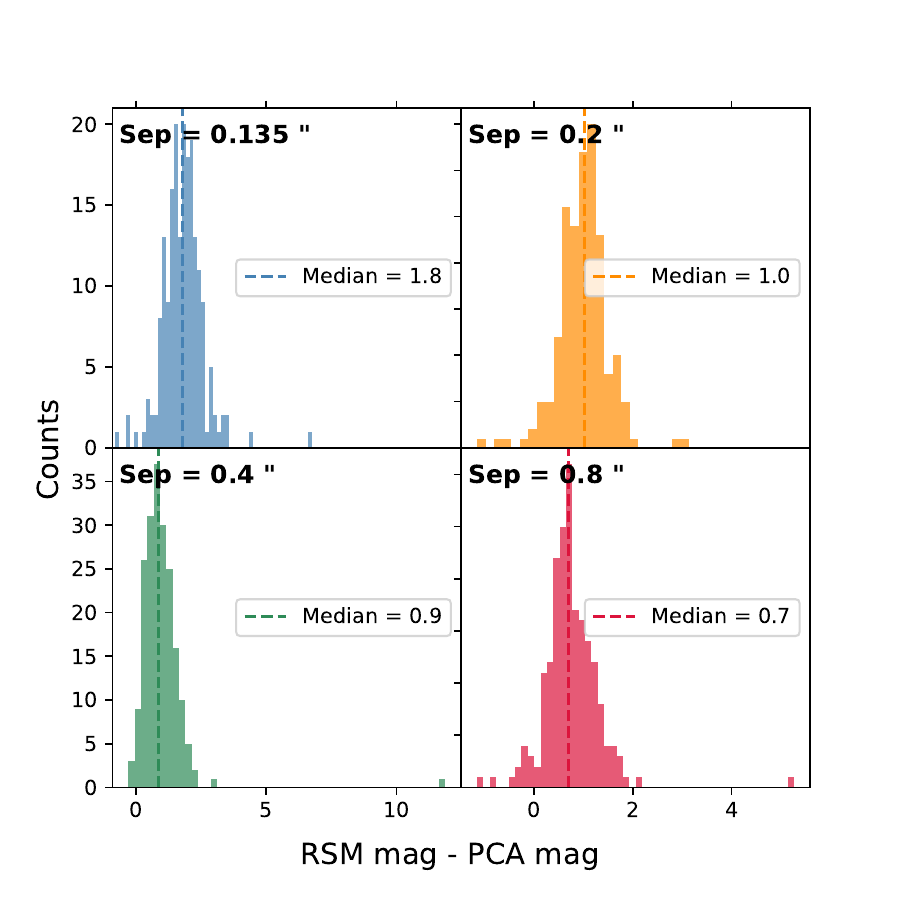}
    \caption{Comparison of RSM and full-frame PCA improvements in magnitude at separations of 0\farcs135, 0\farcs2, 0\farcs4, and 0\farcs8. The median difference is shown as a dashed line, while histograms represent the full dataset.}
    \label{PCA_RSM_mag}
\end{figure}

This improvement is further highlighted in Fig.~\ref{PCA_RSM_mag}, where the magnitude difference between RSM and full-frame PCA sensitivity limits shows a median improvement of 1.8~mag at 0\farcs135, emphasizing RSM's superior detection at small separations and its ability to distinguish planetary signals amidst stellar residual noise. The improvement closely aligns with that achieved using PACO \citep{2023A&A...675A.205C}, with both methods delivering an improvement of approximately a factor of five at small separations and a factor of about two overall, compared to full-frame PCA. These results emphasize the consistent advancements in sensitivity limits across different post-processing techniques. Figure~\ref{PCA_RSM_mag} shows that a few datasets exhibit slightly better performance with PCA than with RSM. This occurs primarily for observations located near the boundaries of their respective clusters, where the adopted RSM parameters may deviate from their true local optima, or in cases where the generalized noise model across the cluster is more restrictive than the dataset’s actual RSM noise properties. These exceptions are rare and do not affect the overall improvement trends observed across the survey, although they do highlight certain limitations of parameter and noise generalization for a small subset of datasets.

\begin{figure*}[t]
    \centering
    \sidecaption
    \includegraphics[width=12cm]{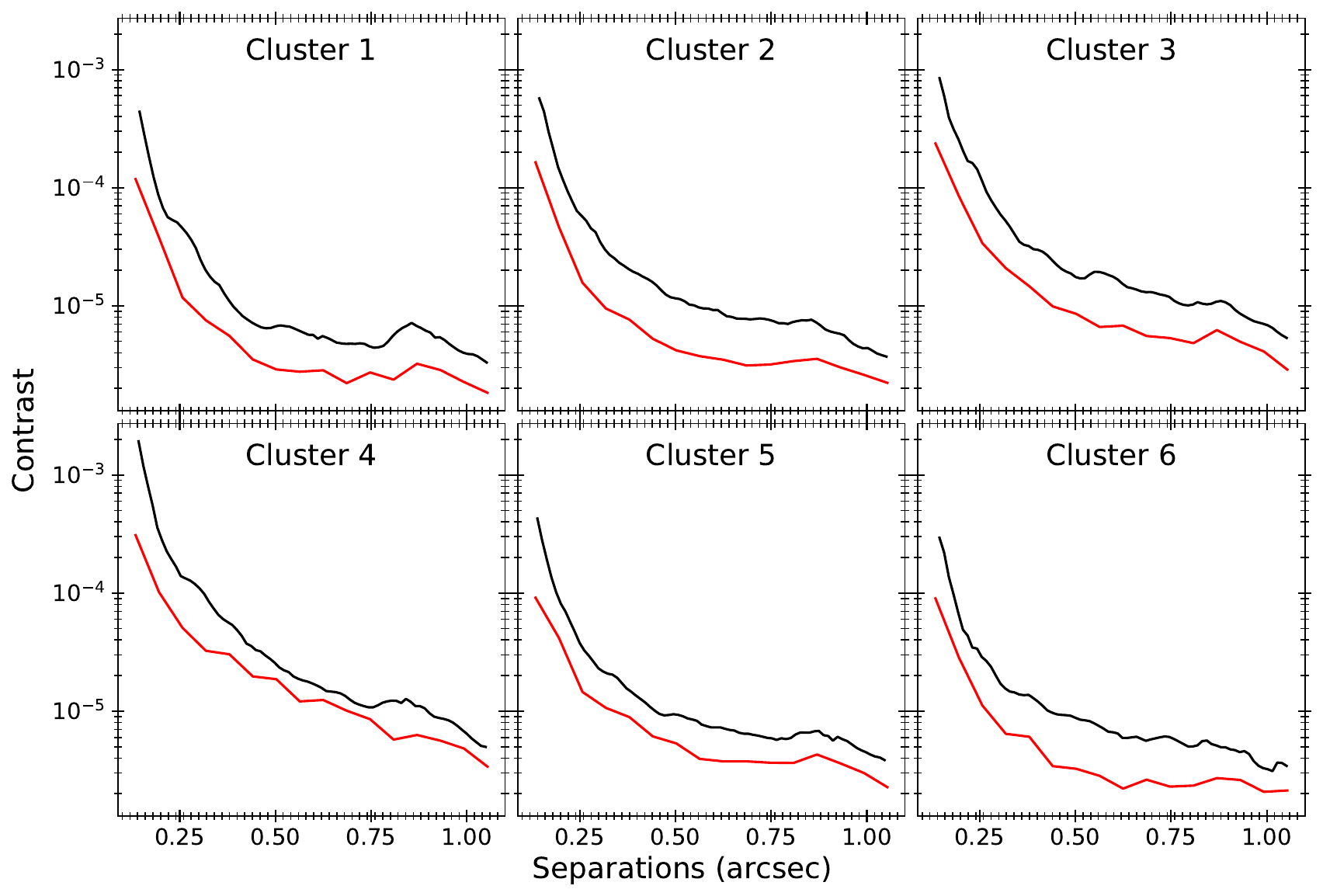}
    \caption{Contrast performance improvements of RSM across various observing conditions in the SHINE survey clusters: median contrast curves are shown in red for RSM and black for PCA, demonstrating RSM's superior efficiency overall.}
    \label{contrasts_different_clusters}
\end{figure*}

This study also allows us to evaluate RSM’s performance under varying observational conditions.  Figure \ref{contrasts_different_clusters} highlights the improvements across distinct clusters in the SHINE survey, demonstrating consistent enhancements across different datasets.
This underscores RSM's robustness in handling diverse observational scenarios and its ability to improve point source detection across the entire survey. 

\section{Identification of point sources}
\label{section_identification_point_sources}

The application of the RSM algorithm to the F150 SHINE sample resulted in the identification of 87 signals across 59 different observations, while 154 observations from the F150 sample showed no detectable signal within the $1\farcs1$ field of view. As previously described, a signal is considered significant when it lies above the detection threshold in all four RSM detection maps, considering the two selected algorithm combinations and the parallel observations performed by SPHERE-IRDIS in the H2 and H3 filters. A dedicated tool was developed to cross-match detections between maps within half a FWHM, ensuring consistent identification of point sources across filter combinations. The photometric and astrometric measurements were derived following the procedure detailed in \citet{2022A&A...666A..33D}. The performance of these measurements was assessed by \citep{2024SPIE13097E..13C} during the Exoplanet Imaging Data Challenge (phase II); this analysis is not discussed further here, as it lies beyond the scope of the present work.

Among the detected signals, 49 were already recovered by \citet{2025A&A...697A..99C}, while 38 were not identified with PACO and required a dedicated analysis to distinguish genuine astrophysical sources from false positives. This counting of false positives does not include some artifacts that tend to appear in RSM maps in the presence of bright companions, at the same angular distance as those companions. This section presents a detailed comparison between the signals detected with RSM and PACO in Sect.~\ref{subsection_comparison_with_PACO}, followed by an analysis of the new point sources recovered exclusively with RSM in Sect.~\ref{subsection_identification_new_point_sources} using proper motion and color-magnitude diagram tests.

\subsection{Comparison with PACO on the F150 sample}
\label{subsection_comparison_with_PACO}

Given the comparable performance of RSM and PACO, a direct comparison of their detected signals provides valuable insight into the relative sensitivity of both methods, the reliability of faint detections, and potential missed detections on either side.
To ensure a fair comparison between PACO and RSM, only the common observations in the archival point-source detections reported in \citet{2025A&A...697A..99C} were considered. These correspond to common H2/H3-band observations covering fields of view between 0\farcs11 and 1\farcs1. Within this sample, PACO reported no detection in 181 observations. Among these, 29 revealed a detection in RSM, while 152 are classified as non-detections in both methods. These 29 observations contain 38 signal detections that are examined in more detail in the next section. In addition to these 181 observations classified as non-detections, PACO recovered 52 signals in the remaining 32 observations, including 24 confirmed planets (observed at different epochs), 24 background stars, and four ambiguous detections. Here, we compare these 52 PACO detections with our RSM results.

\begin{figure}[t]
    \centering
    \includegraphics[width=0.5\textwidth]{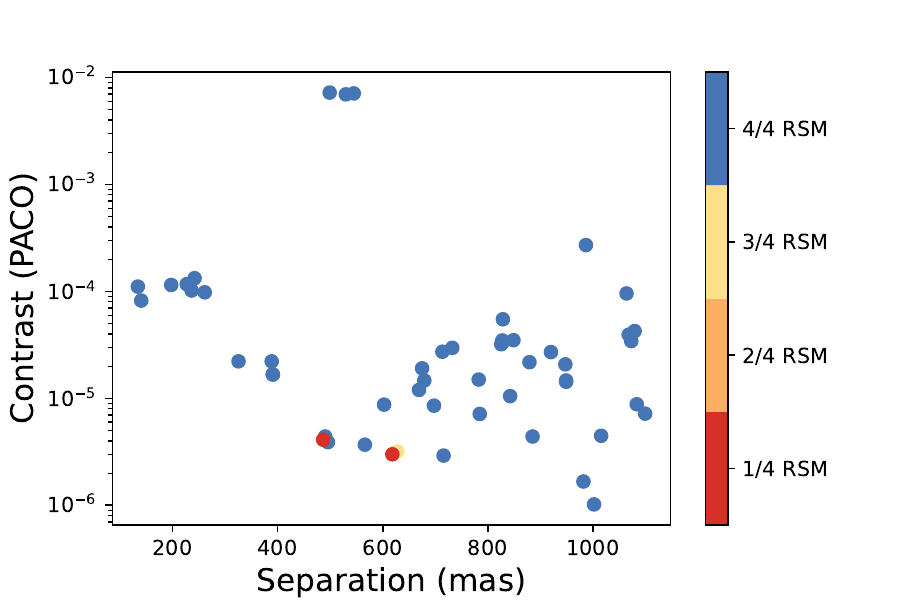}
    \caption{Contrast distribution and separation of the 52 PACO-detected signals and their recovery by RSM. Dark blue points represent signals recovered in all four RSM maps, other colors correspond to sources not recovered by RSM, or recovered in only some RSM maps.}
    \label{figure_signals_contrast_PACO_RSM}
\end{figure}

Out of the 52 PACO detections, 49 were also recovered by RSM based on our four-map detection criterion. Figure~\ref{figure_signals_contrast_PACO_RSM} presents the contrast values of each PACO-detected signal, with colors indicating the recovery status in RSM; dark blue points represent signals fully recovered in four maps, while other colors correspond to unrecovered sources. As shown in the figure, most signals detected by PACO were also recovered by RSM, with only a few exceptions.

The unrecovered signals correspond to the confirmed planet HD~95086~b (in two observations), and a background star around HD~151726, although these sources were recovered in at least one of the RSM maps. The primary reason for this discrepancy lies in the different detection criteria employed by the two algorithms. The RSM method requires recovery above the detection threshold in both H2 and H3 filters, whereas PACO derives a single signal-to-noise ratio (S/N) that combines information from both spectral channels. Consequently, sources with marginal S/N in one band but a stronger signal in the other may be recovered by PACO but not by RSM.

This behavior is clearly illustrated by the case of HD~95086~b, observed on 2015-05-05 and 2015-05-11, with PACO S/N values of 6.8 and 5.0, respectively. \citet{2022A&A...664A.139D} showed that the planet-star contrast for this target is twice as large in H3 compared to H2 (see their Fig.~5), consistent with the \citet{2025A&A...697A..99C} database, where the H2 contrast errors are nearly as large as the measured contrast values themselves.

\subsection{Identification of new point sources}
\label{subsection_identification_new_point_sources}

The application of the RSM algorithm to the F150 SHINE sample revealed 38 newly detected point sources across 29 different observations for 27 stars. Among these targets, ten stars have additional epochs within the F150 sample, while the remaining 17 correspond to single-epoch observations.

Multi-epoch data enable the differentiation between false positives and genuine astrophysical sources through proper motion tests.
Out of the ten targets that have additional epochs, two exhibit detections in more than one epoch, while the remaining eight show detections in only a single epoch despite the availability of multiple observations.

For detections found in only one epoch among several available observations, both the expected position of the source within the field of view (FOV) and the observing conditions of the non-detection epochs were examined. If the expected position in other epochs laid outside the FOV or if the observing conditions were less favorable, these signals were retained for further analysis; otherwise, they were classified as false positives. This resulted in one detection being retained for further analysis, while 12 signals were classified as false positives. These false positives are primarily associated with observations exhibiting low Strehl ratios and limited field rotation.

Regarding the multi-epoch detections, RSM successfully recovered PZ~Tel in an observation from 2014, which was not detected by PACO, unlike other epochs where the signal was identified by both methods. RSM also recovered a background star around HD~164249 identified through the proper motion test between the 2015 and 2016 epochs. These signals were not reported by \citet{2025A&A...697A..99C} but were previously mentioned by \citet{2021A&A...651A..71L}.

\begin{figure}[t]
\centering
\includegraphics[width=0.5\textwidth]{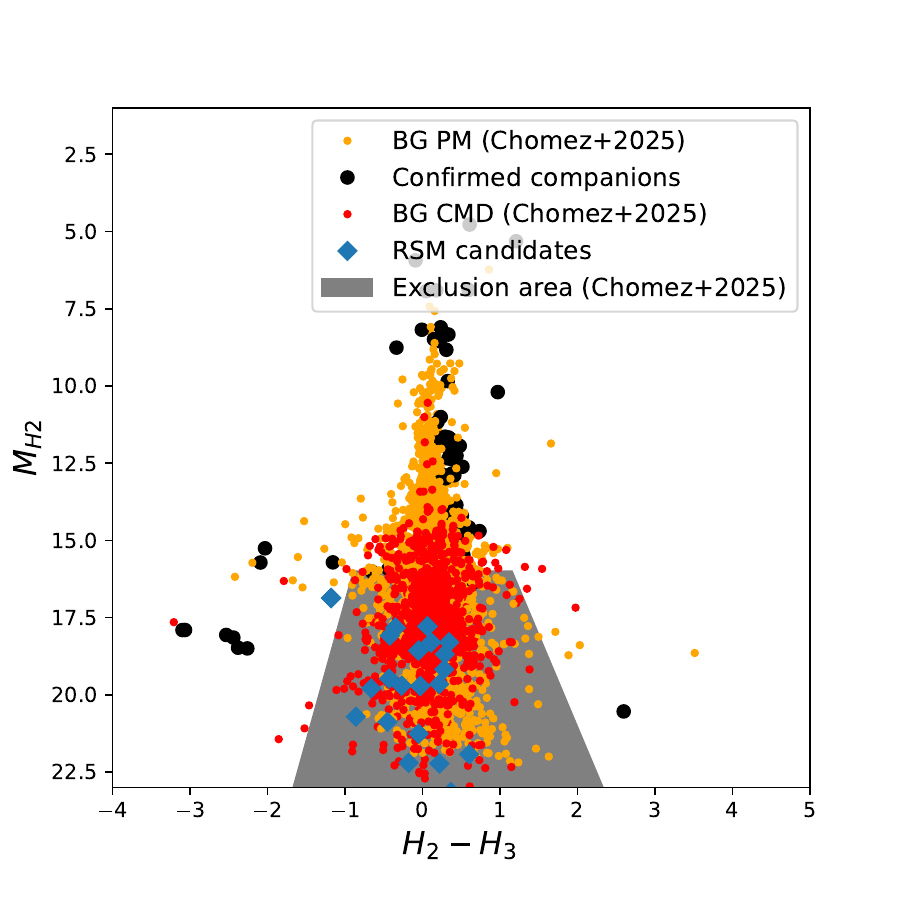}
\caption{Color–magnitude diagram showing the new RSM detections in the F150 SHINE sample, plotted alongside the signals reported in \citet{2025A&A...697A..99C} for confirmed exoplanets, background stars identified through proper motion, and background stars classified via CMD analysis. The gray shaded area indicates the exclusion region for background stars defined by \citet{2025A&A...697A..99C}.}
\label{figure_color_magnitude_diagram}
\end{figure}

Verifying the nature of a signal for single-epoch targets is more challenging. In these cases, following \citet{2025A&A...697A..99C}, a color-magnitude diagram (CMD) analysis is employed to assess the likelihood of each detection being consistent with a planetary companion or a background star. The 17 targets that had no additional epochs within the SHINE F150 survey generated a total of 22 significant signals in our RSM maps. To these 22 signals, we add the ambiguous candidate identified in the previous paragraph, and perform a CMD analysis to evaluate the astrophysical relevance of each signal.

The CMD is shown in Fig.~\ref{figure_color_magnitude_diagram} for all the detections reported in \citet{2025A&A...697A..99C}, supplemented by the new RSM candidates.
As discussed by \citet{2025A&A...697A..99C}, background stars occupy a well-defined region in the diagram, and we adopt their background-star region, referred to as the exclusion area in Fig.~\ref{figure_color_magnitude_diagram}.

Most of the RSM candidates fall within this exclusion region, indicating that they are likely background stars. Two signals, falling outside the plot range of Fig.~\ref{figure_color_magnitude_diagram}, display photometric properties inconsistent with an astrophysical source and are therefore classified as false positives. One last source lies just outside the exclusion region. It exhibits photometric characteristics consistent with an exoplanet-like nature and is thus considered a promising planet candidate. Appendix~\ref{appendix_candidates} presents several examples of newly identified background-star detections and compares their corresponding PCA signal-to-noise ratios with the RSM scores.

\begin{figure}[t]
    \centering
    \includegraphics[width=0.5\textwidth]{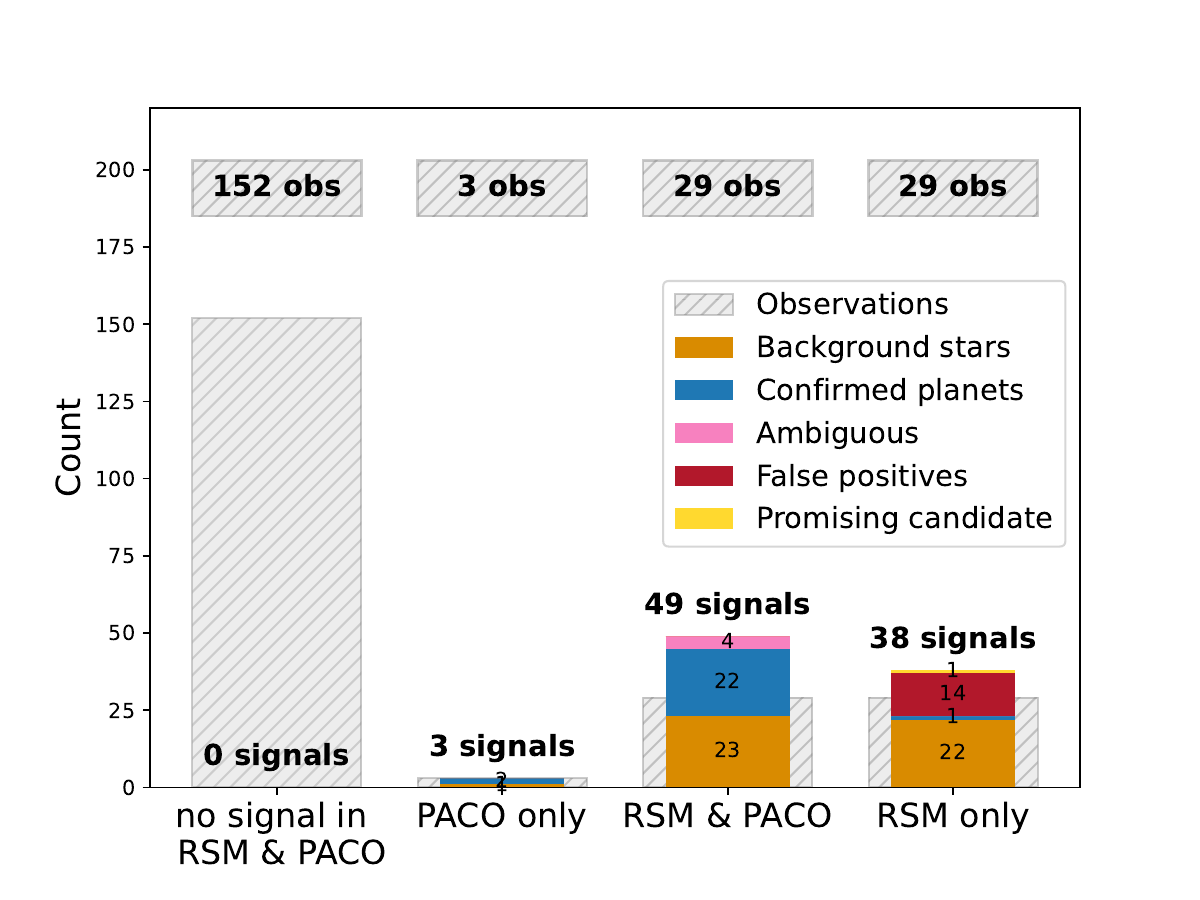}
    \caption{Summary of the detection statistics for the F150 sample, comparing the outcomes of RSM and PACO across four cases: no signal in either method, signal in PACO only, signal recovered by both methods, and signal recovered by RSM only. The dashed gray bars indicate the number of observations in each category, with the corresponding counts labeled above them. The colored stacked bars represent the number of detected signals, with colors denoting their classification: confirmed planets, background stars, ambiguous sources, false positives, and the newly identified promising candidate.}
    \label{figure_numbers}
\end{figure}

Figure~\ref{figure_numbers} summarizes the detection statistics for the F150 sample, including the number of observations showing no signal, the signals recovered by RSM, by PACO, or by both methods, and the classification of each detected source. The figure also serves as a comprehensive visual summary of all numerical values discussed in the paper.

\section{Conclusions}
This study introduces new methods for determining detection thresholds in high-contrast imaging, utilizing the RSM applied to the SHINE F150 survey. Clustering based on simple environmental parameters using k-means effectively classified observations into meaningful groups, such as those with large number of frames, poor conditions, or specific effects such as wind driven halo and low-wind effect. Spearman correlation tests confirmed the distinct noise characteristics in each cluster and validated the cluster representatives in terms of environmental and noise parameters. We have additionally highlighted how observational conditions influence RSM noise distribution in final RSM maps. By fitting a lognormal distribution at each separation, detection thresholds with a $3 \times 10^{-7}$ FAP revealed notable variations across different clusters, emphasizing the importance of adapting thresholds to observational conditions. We also compared lognormal thresholds with those derived from maximizing the F1 score. While both approaches are compatible, the F1 score method better addresses individual bright false positives, making it suitable for single observations. Conversely, the lognormal fit reflects the overall noise behavior and provides a more generalizable solution for large surveys.

Leveraging the diverse noise behavior across frames in the science cubes and applying different PSF subtraction techniques further enhanced RSM's ability to distinguish planetary signals from speckles. This led to narrower RSM noise distributions, reduced thresholds, and improved detection limits, achieving a fivefold improvement at 135 mas and a twofold overall enhancement at larger separations compared to \citet{2021A&A...651A..71L} for the F150 SHINE sample. Compared to \citet{2025A&A...697A..99C}, who used the PACO algorithm, applying RSM to the F150 sample achieved comparable performance. Our study led to the identification of 87 signals, 49 of which were previously reported by \citet{2025A&A...697A..99C}. Among the 38 new detections, 22 were classified as background stars, 14 as false positives, one turned out to be a known low-mass companion, and one was identified as a new exoplanet candidate for which follow-up observations are needed. This result highlights RSM’s robustness and potential for further applications.

Finally, the methodology developed in this work can be directly extended to the full SHINE survey and adapted to other ground-based high-contrast imaging programs, such as the Gemini-GPIES \citep{2020AJ....159...71N}, SPHERE-BEAST\citep{2019A&A...626A..99J,2021A&A...646A.164J}, or NACO-ISPY \citep{2020A&A...635A.162L} surveys. This framework provides a robust and versatile foundation for statistically consistent planet detection and performance characterization across a wide range of observing conditions.

\section*{Data availability}
\label{data_availability}

The RSM algorithm is publicly available on GitHub as a Python package\footnote{\url{https://github.com/chdahlqvist/RSMmap}}.
The raw and reduced data used in this study can be requested through the High-Contrast Data Centre (HC-DC)\footnote{\url{https://hc-dc.cnrs.fr}}.
The code developed for the analysis presented in this work is openly accessible on GitHub\footnote{\url{https://github.com/marisabalbal/RSM_f150SHINE}}.
The full point-source candidate list will be presented in a forthcoming publication, together with an NA-SODINN analysis of the SHINE F150 sample (Cantero et al., in prep). The table for the 213 F150 SHINE observations used in this study, along with their observing conditions, is only available in electronic form at the CDS via anonymous ftp to cdsarc.u-strasbg.fr (130.79.128.5) or via \url{http://cdsweb.u-strasbg.fr/cgi-bin/qcat?J/A+A/}.

\begin{acknowledgements}
    This work has made use of the High Contrast Data Centre, jointly operated by OSUG/IPAG (Grenoble), PYTHEAS/LAM/CeSAM (Marseille), OCA/Lagrange (Nice), Observatoire de Paris/LESIA (Paris), and Observatoire de Lyon/CRAL, and supported by a grant from Labex OSUG@2020 (Investissements d’avenir – ANR10 LABX56).
    This project received funding from the European Research Council (ERC) under the European Union’s Horizon 2020 research and innovation program (grant agreement No 819155). Computational resources were provided by the Consortium des Équipements de Calcul Intensif (CÉCI), funded by the Fonds de la Recherche Scientifique de Belgique (F.R.S.-FNRS) under Grant No. 2.5020.11 and the Walloon Region. \\We extend our gratitude to Faustine Cantalloube, Valentin Christiaens, and Sandrine Juillard for their insightful discussions.  
\end{acknowledgements}
\bibliographystyle{aa}
\bibliography{references}

\appendix

\twocolumn

\section{Pixel correlation between datasets}
\label{appendix_correlations}
To assess how well the chosen cluster centers reflect the noise behavior of other datasets within the same cluster, we performed a pixel correlation test. This test computes the correlation between the selected cluster center and the other datasets in the cluster. Additionally, we compare this correlation to that of the dataset exhibiting the highest correlation within the cluster. 

\begin{figure}[t]
    \centering
    \includegraphics[width=0.5\textwidth]{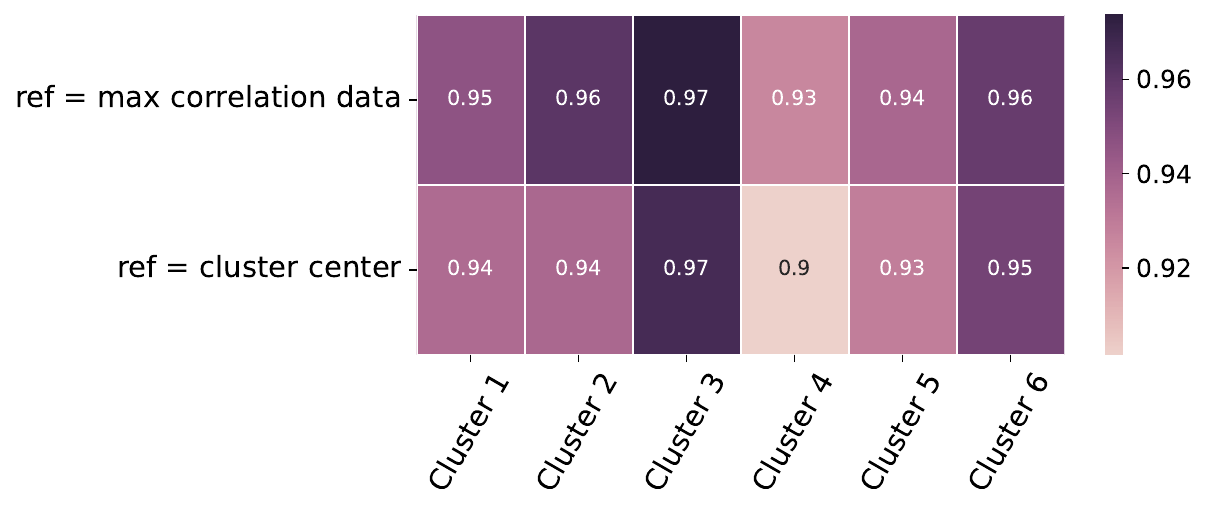}
    \caption{Spearman rank correlation comparison between cluster centers and most correlated datasets in each cluster.}
    \label{correlation_centers}
\end{figure}

To study the correlation between datasets, we adopted the approach used in \citet{2024A&A...688A.185J}, using a correlation metric from the VIP package based on the Spearman rank correlation coefficient. This coefficient was chosen to avoid the normality assumption required by the Pearson coefficient and to circumvent the dependence of the Structural Similarity Index (SSIM) on the image dynamic range, which will be normalized later in the post-processing.  For each dataset, we identified the frame within the science cube that exhibits the highest correlation with other frames, designating it as the representative frame of that dataset. We then computed the pixel correlation between this representative frame and the representative frames of all other datasets in the cluster. This process was repeated for all datasets within one cluster as well as for all clusters. The dataset exhibiting the highest correlation with the other datasets in its cluster is expected to share the most prominent noise features and thus be the most representative. We then compared the Spearman correlation of the selected cluster centers, chosen based on environmental conditions, to that of the most correlated dataset in each cluster. The results are shown in Figure \ref{correlation_centers}.

The high correlation values for the cluster centers confirm that these selections reliably represent their respective clusters. This correlation metric proves less effective for clusters affected by strong wind-driven halo effects (clusters 4 and 5). This limitation arises from the rotation of wind-driven speckles with the parallactic angle \citep{2020A&A...638A..98C}, which introduces additional complexity. In such cases, high correlation coefficients tend to identify datasets with less pronounced wind effects rather than true feature similarity, reducing the effectiveness of correlation as a metric in these specific scenarios.

\section{Clustering illustration using principal components}
\label{appendix_clustering}
\begin{figure}[t]
    \centering
    \includegraphics[width=\linewidth]{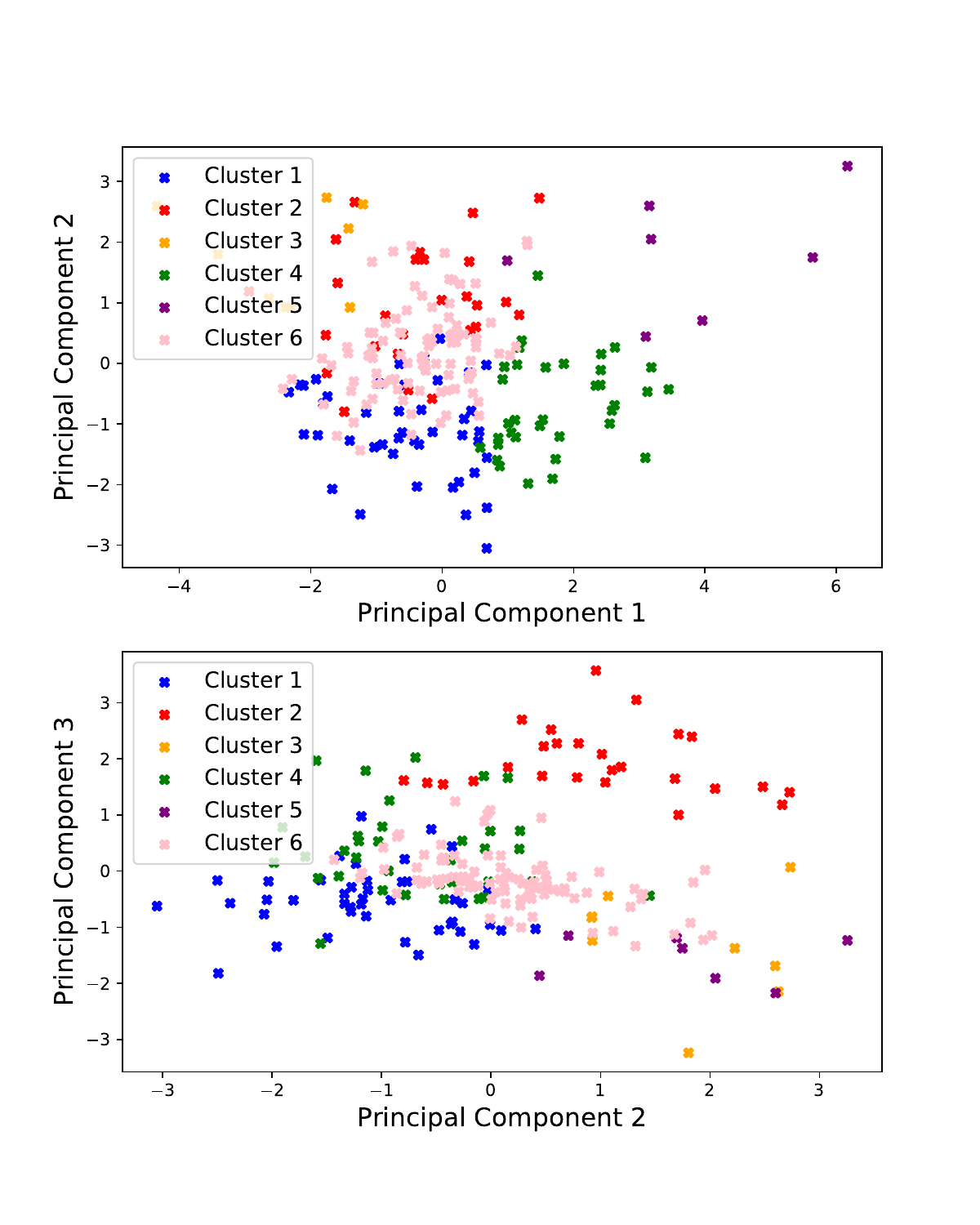}
    \caption{Visualization of clusters using principal component analysis. \textit{Top:} First versus second component projection. \textit{Bottom:} Second versus third component projection.}
    \label{clustering_PC}
\end{figure}

To analyze the distinctions between clusters, we projected the environmental datasets onto their first three principal components. Figure~\ref{clustering_PC} (top) shows that most clusters are well-separated, although clusters two and 6 overlap. This overlap is resolved when examining the third principal component, as shown in Plot \ref{clustering_PC} (bottom).

\section{Optimal parameters for cluster centers}
\label{table_optimal_parameters_RSM}

\begin{table*}

\caption{Optimal parameters for PSF subtraction techniques and RSM for each cluster center.} 
\label{tab:RSMparams}
\centering
\begin{tabular}{lcccccc}
\hline\hline
Parameters & Center 1 & 
Center 2 & Center 3 & Center 4 & Center 5 & Center 6
\\ [0.5ex] 
\hline

APCA components & 20 & 14 & 18 & 8 & 7 & 12  \\

APCA segments & 4 & 3 & 3 & 3 & 3 & 4  \\

APCA FOV rotation & 0.42 & 0.29 & 0.26 & 0.31 & 0.5 & 0.76  \\

NMF components & 9 & 10 & 10 & 22 & 11 & 6  \\

LOCI tolerance & 0.0038 & 0.0016 & 0.0012 & 0.0057 & 0.0096 & 0.002  \\

LOCI FOV rotation & 0.49 & 0.14 & 0.48 & 0.1 & 0.13 & 0.73  \\

FM KLIP components & 9 & 19 & 8 & 12 & 5 & 5  \\

FM KLIP FOV rotation & 0.51 & 0.27 & 0.27 & 0.34 & 0.25 & 0.81  \\

FM LOCI tolerance & 0.0042 & 0.0074 & 0.0072 & 0.0073 & 0.0099 & 0.0016  \\

FM LOCI FOV rotation & 0.11 & 0.13 & 0.13 & 0.11 & 0.2 & 0.61  \\

APCA $\delta$ & 5 & 5 & 5 & 5 & 5 & 5  \\

NMF $\delta$ & 5 & 5 & 5 & 5 & 5 & 5  \\

LOCI $\delta$ & 5 & 5 & 5 & 5 & 5 & 5  \\

FM KLIP $\delta$ & 5 & 5 & 5 & 5 & 5 & 5  \\

FM LOCI $\delta$ & 5 & 5 & 5 & 5 & 5 & 5  \\

APCA crop size & 4 & 4 & 4 & 4 & 4 & 4  \\

NMF crop size & 4 & 4 & 4 & 4 & 4 & 4  \\

LOCI crop size & 4 & 4 & 4 & 4 & 4 & 4  \\

FM KLIP crop size & 7 & 7 & 7 & 7 & 7 & 7  \\

FM LOCI crop size & 7 & 7 & 7 & 7 & 7 & 7  \\

APCA variance & SM & ST & SM & SM & SM & FR  \\

NMF variance & SM & SM & SM & SM & SM & SM  \\

LOCI variance & FR & ST & ST & SM & SM & ST  \\

FM KLIP variance & FR & ST & ST & SM & ST & SM  \\

FM LOCI variance & SM & SM & ST & ST & ST & ST  \\
\hline
\end{tabular}

\tablefoot{
The RSM parameters include the planetary flux multiplicative factor ($\delta$), defined as a multiple of the standard deviation, the crop size, and the variance term used to determine the noise estimation region within the annulus. The noise estimation methods are denoted as follows: "SM" (Segment with mask-based estimation), "ST" (Spatio-Temporal estimation), and "FR" (Frame-based estimation) \citep[for further details, see][]{2021A&A...656A..54D}. }
\end{table*}

\section{Variation of optimal RSM parameters across cluster centers and individual datasets}
\label{appendix_similarities_RSM}
\begin{figure}[t]
    \centering
    \includegraphics[width=0.5\textwidth]{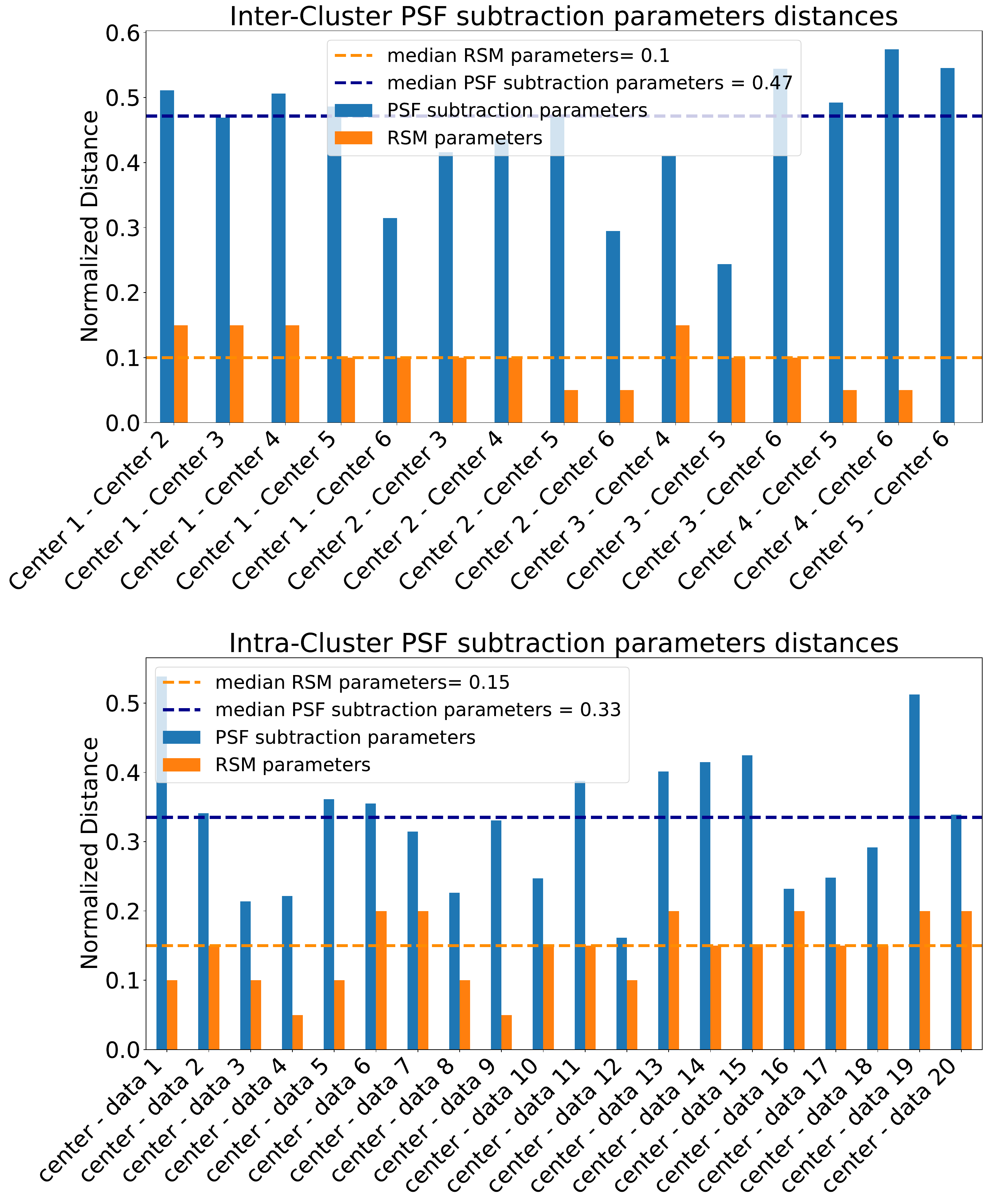}
    \caption{Normalized distances for PSF subtraction (blue) and RSM parameters (orange) between cluster centers (top) and datasets within a single cluster (bottom). }
    \label{similarities}
\end{figure}

This section examines the impact of clustering on optimal parameter selection in RSM, emphasizing how different observing conditions influence the choice of PSF subtraction technique parameters and assessing the reliability of generalizing cluster center parameters to all datasets within a cluster. To quantify these effects, we measure the normalized distance between parameters. For numerical parameters, we compute their normalized distances, while for categorical parameters (e.g., the noise estimation methods), we use the dissimilarity index, a normalized measure of how different two RSM parameter sets are, with 0 indicating identical configurations and 1 indicating distinct ones. The mean normalized distance across all parameters is then calculated for each cluster center pair. Figure~\ref{similarities} (top) presents these distances, revealing substantial variability in PSF subtraction parameters, with a median distance of 0.47, while RSM parameters exhibit lower variability, with a median distance of 0.1. This indicates significant differences in optimal parameter choices between cluster centers for PSF subtraction techniques. Overall, different PSF subtraction methods exhibit varying median distances across the clusters, with APCA having the lowest median distance of 0.33 and LOCI the highest at 0.68.

To assess the validity of generalizing parameters from a single representative dataset to an entire cluster, we conducted the same analysis within a small but diverse cluster (Cluster 1). Figure~\ref{similarities} (bottom) illustrates the normalized distances between 20 datasets and the cluster center, showing a reduced median distance of 0.33 for PSF subtraction parameters. One dataset (HD~109573) failed to converge during optimization due to the presence of bright disk emission, explaining the use of 20 instead of 21 datasets.  The slightly larger distance of 0.15 observed for RSM parameters primarily stems from variations in the noise estimation region parameter, which has minimal impact on the final RSM maps. In contrast, PSF subtraction parameters have a much stronger influence on the results. The reduction in parameter distance suggests that datasets with similar observing conditions tend to converge toward more comparable parameter choices.

Overall, this analysis underscores the substantial differences in optimal parameter selection between different cluster centers, highlighting the need to adapt parameter choices to specific datasets across a survey. At the same time, it demonstrates that datasets with similar observing conditions generally exhibit closer parameter distances. However, some datasets within a single cluster still show notable variations, reflecting dataset-specific noise properties. Despite these differences, we adopt a cluster-based parameter optimization approach in this study, as optimizing RSM parameters for each individual dataset remains computationally costly.

\section{Comparison with the Dahlqvist contrast curve definition}
\label{appendix_completeness}
\begin{figure}[t]
    \centering
    \begin{subfigure}[b]{.5\textwidth}
        \centering
        \includegraphics[width=\textwidth]{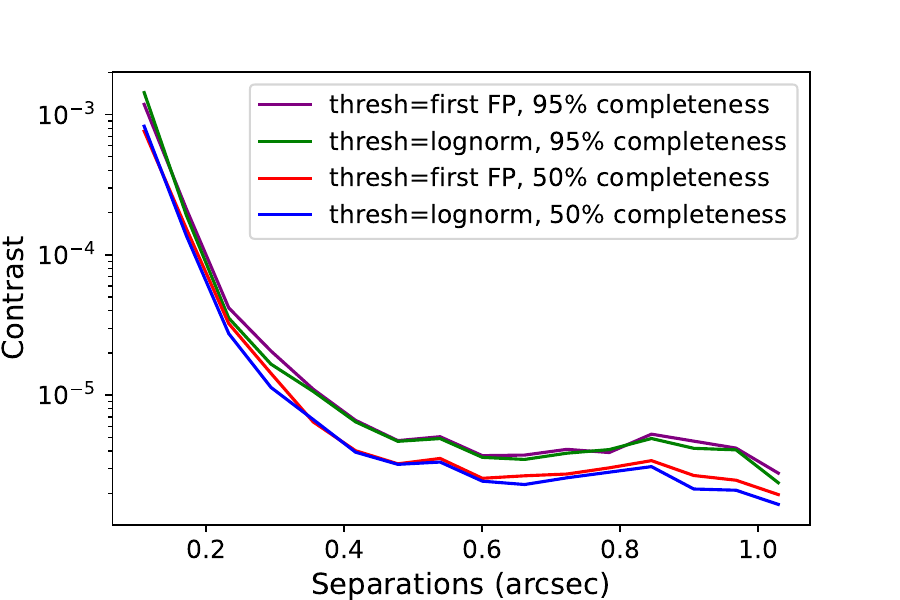}  
        
        \caption{}
        \label{completeness_center}
    \end{subfigure}
    \begin{subfigure}[b]{.5\textwidth}
        \centering
        \includegraphics[width=\textwidth]{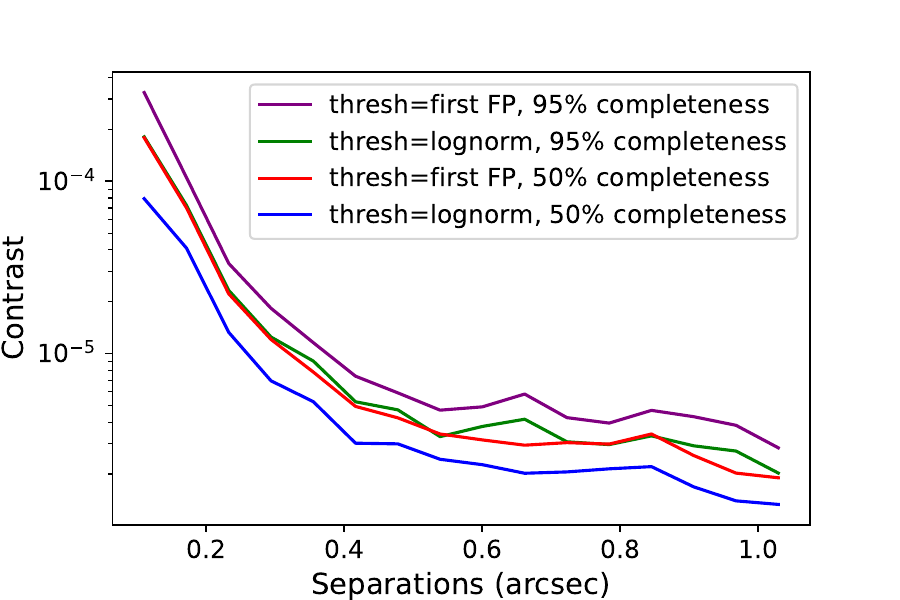}  
        
        \caption{}
        \label{completeness_target}
    \end{subfigure}
    \caption{Comparison of contrast curves for two datasets, one with no potential companion (a) and one with a potential companion (b). Contrasts are generated with
    95\% completeness using the first false positive threshold and lognormal threshold (in purple and green), and 50\% completeness for both thresholds (in red and blue)}
    \label{completeness}
\end{figure}

This section compares the contrast curve methodology used in \citet{2022A&A...666A..33D} for the SHARDDS survey with the approach applied in our study. \citet{2022A&A...666A..33D} define the detection threshold based on the first false positive within the field of view and considers 95\% recovery of injected signals. In contrast, our approach employs a detection threshold corresponding to an FAP of $3 \times 10^{-7}$ under a lognormal distribution at a given separation and evaluates contrast at the 50\% injection recovery level. Figure \ref{completeness} presents these contrast curves for two datasets: one without a potential signal (Fig.~\ref{completeness_center}) and another with a potential signal (Fig.~\ref{completeness_target}). In Fig.~\ref{completeness_center}, the contrast curves derived from different detection thresholds remain consistent at the same completeness levels. However, this trend does not hold in Fig.~\ref{completeness_target}, where the contrast curve based on the lognormal detection threshold outperforms that derived from the first false positive at both completeness levels.  In both figures, there is only a minor gap between the contrast levels required for 50\% and 95\% signal recovery. This result highlights a key limitation of the approach in \citet{2022A&A...666A..33D}, where the contrast is influenced by the brightest signal in the field of view without determining whether it originates from noise or a potential candidate —a limitation that also applies to thresholds based on the maximum F1 score. Moreover, this method defines the detection threshold based on the entire field of view rather than at each separation, limiting its ability to assess contrast variations across different separations.

\section{Comparing RSM noise distribution models}
\label{appendix_lognormal_distribution}
\begin{figure}[t]
    \centering
    \includegraphics[width=0.5\textwidth]{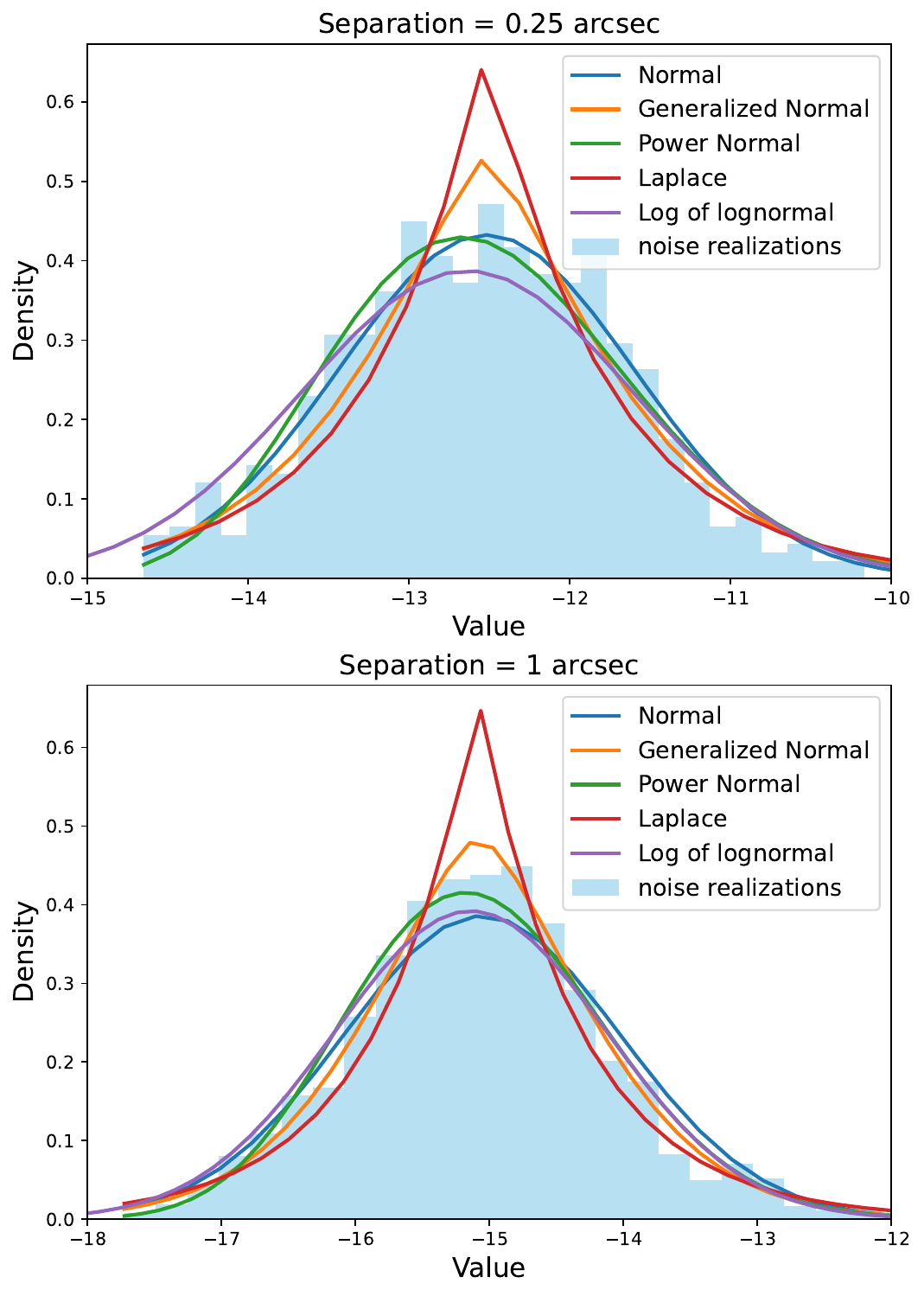}
    \caption{Histograms of independent RSM noise realizations (log scale) in the RSM maps of Cluster 1, obtained using APCA RSM, at 0\farcs25 (top) and at 1\arcsec\ (bottom). The fitted distributions include normal (blue), generalized normal (orange), power-normal (green), Laplace (red), and log of lognormal (purple).}
    \label{histograms}
\end{figure}

The distribution of independent RSM noise realizations, as discussed in Section~\ref{subsection_lognormal_distribution}, exhibits a near-normal profile when viewed on a logarithmic scale. Here, we investigate various normal-like distributions to determine the best fit for the datasets. Specifically, we compare the log of lognormal distribution (three parameters), the normal distribution (two parameters), the power-normal distribution (three parameters), the generalized normal distribution (three parameters), along with the Laplace distribution (two parameters). Figure~\ref{histograms} presents the distribution of independent RSM noise realizations using APCA in the RSM context for cluster 1 at two separations, along with the fitted distributions. It is important to note that the data used in this section fall below the $3 \times 10^{-7}$ threshold. The histograms demonstrate a clear normal-like profile, with all tested distributions providing a reasonable fit, with the laplace being the least.

\begin{table}   
\centering                                    
\caption{Kolmogorov-Smirnov test for different distributions fits for cluster 1 RSM noise realization samples: p-value at 0\farcs25 and 1\arcsec using RSM APCA.}
\begin{tabular}{c c c }         
\hline\hline 
Distributions & sep = 0\farcs25 & sep = 1\arcsec  \\    
\hline                                   
    Log of Lognorm & 0.5 & 0.09 \\      
    Norm & 0.72 & 0.00      \\
    Generalized norm & 0.43 & 0.71\\
    Power norm & 0.56 & 0.17\\
    Laplace & 0.04 & 0.00 \\
\hline                                           
\end{tabular} 
\label{table_KS_1}
\end{table}

To assess the goodness of fit, we perform a Kolmogorov-Smirnov (KS) test and compute the p-values for each distribution at different separations. Table~\ref{table_KS_1} presents the results for Cluster 1, where RSM noise realizations were generated using APCA in the RSM detection map. The table shows that the generalized normal, the logarithm of the lognormal, and the power-normal distributions achieve p-values above 0.05, indicating a good fit. While the generalized normal and power-normal distributions yield high p-values, they do not consistently perform well across all clusters (e.g., failing for Cluster 6). Given its strong overall fit across different clusters and PSF subtraction technique combinations, we adopt the lognormal distribution as the preferred model.

\section{Maximum F1 score threshold generalized across a cluster}
\label{appendix_F1_centers}

\begin{figure}[t]
    \centering
    \centering
    \includegraphics[width=0.45\textwidth]{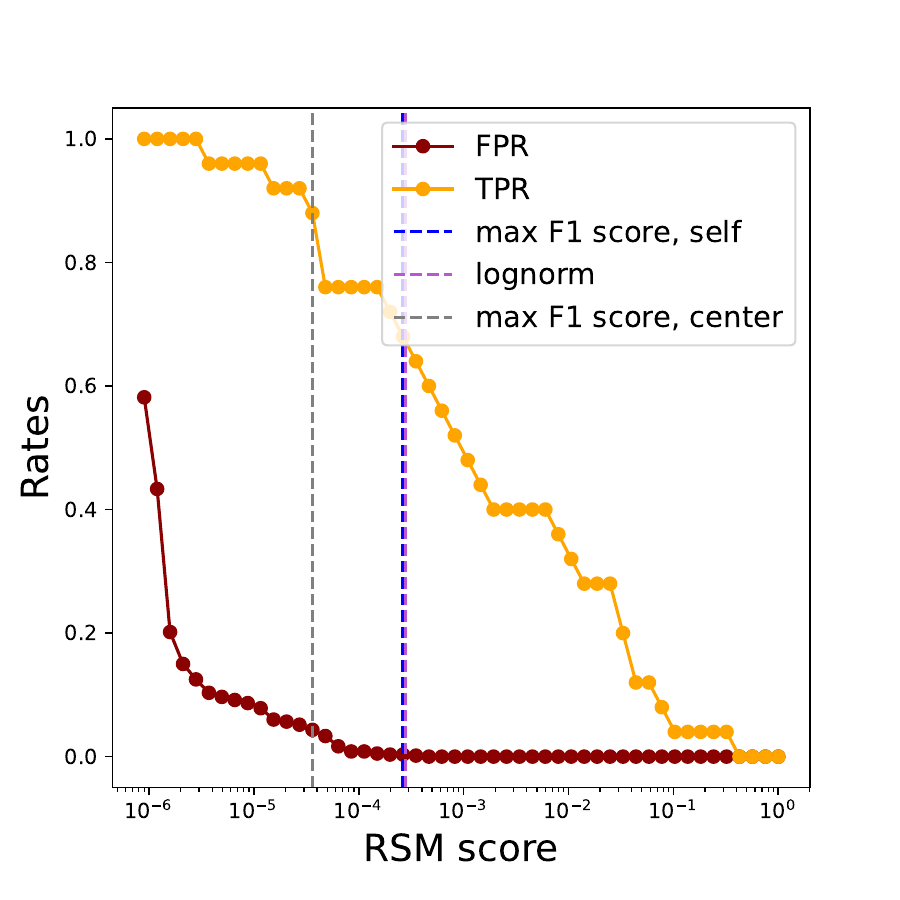}  
    \caption{Maximum F1 score threshold for the cluster center 1 BD-15 705 (in pink) applied on the HIP~92680 dataset, in comparison to the maximum F1 score computed by the dataset HIP~92680 itself denoted as "self" (in cyan) and the lognormal threshold (in dark blue) done by the cluster 1, at separation $0\farcs55$. }
    \label{F1_score_centers}
\end{figure}

This section evaluates the generalization of thresholds determined using the maximum F1 score from a given dataset across other datasets with similar observing conditions. The maximum F1 score is highly sensitive to false positives, often influenced by the brightest speckle in the representative dataset rather than the overall noise profile. Generalizing such thresholds can lead to inconsistencies, as the presence of these speckles reflects specific environmental parameters and their interaction with adaptive optics. These factors are inherently unique to individual observations and cannot be reliably generalized, even under broadly similar observing conditions. Figure \ref{F1_score_centers} illustrates this issue by generalizing the maximum F1 score threshold from the cluster center BD-15 705 to the dataset HIP~92680 in cluster 1. The threshold derived from the cluster center (pink) shows a factor of ten difference compared to the self-computed maximum F1 score ( cyan) and the lognormal threshold (dark blue). This significant discrepancy leads to the generalized threshold accepting false positives that would otherwise be excluded, underscoring the difficulties in applying overly tuned thresholds across datasets.

\section{Comparison between RSM maps and PCA signal-to-noise ratio maps}
\label{appendix_candidates}
This section presents a comparison between the RSM maps and the corresponding PCA signal-to-noise ratio (S/N) maps for several newly identified background-star detections. Specifically, we compare the optimal RSM maps obtained in the H2 and H3 filters using the combined APCA–NMF–LOCI–FMKLIP configuration with the five-component PCA S/N maps.

The distinction between the two approaches is evident: in the PCA S/N maps, many signals are barely distinguishable from the surrounding noise, whereas in the RSM maps these same sources stand out clearly with significantly higher RSM scores. Based on the color–magnitude diagram (CMD) analysis, these signals are classified as background stars.
\begin{figure*}[t]

    \centering
    
    \begin{subfigure}[b]{\textwidth}
        \centering        \includegraphics[width=0.9\textwidth]{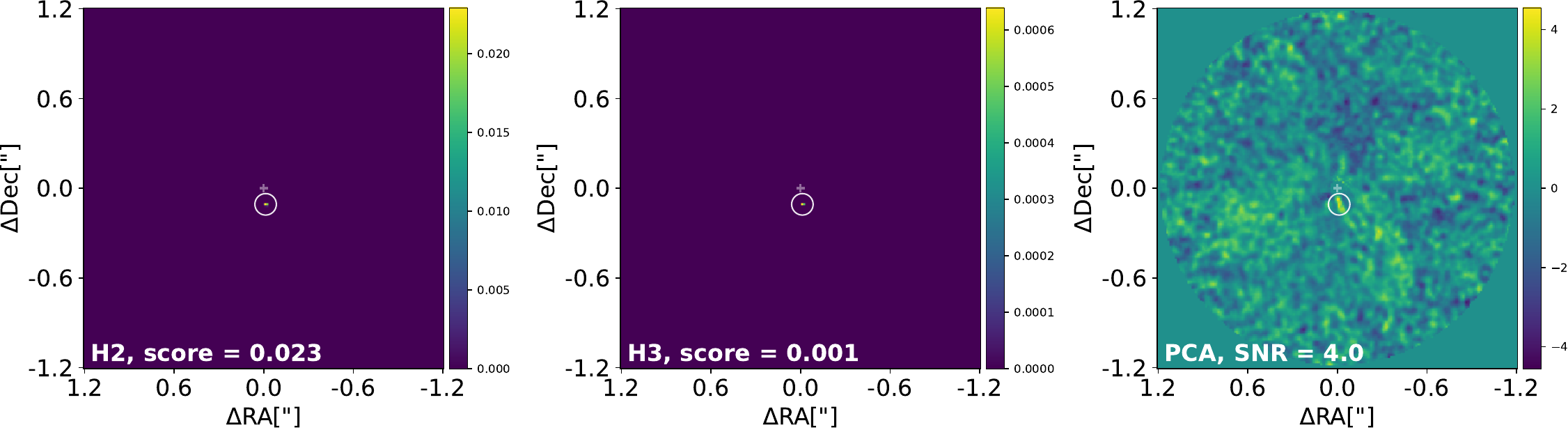}  
    \end{subfigure}
    
    \begin{subfigure}[b]{\textwidth}
        \centering        \includegraphics[width=0.9\textwidth]{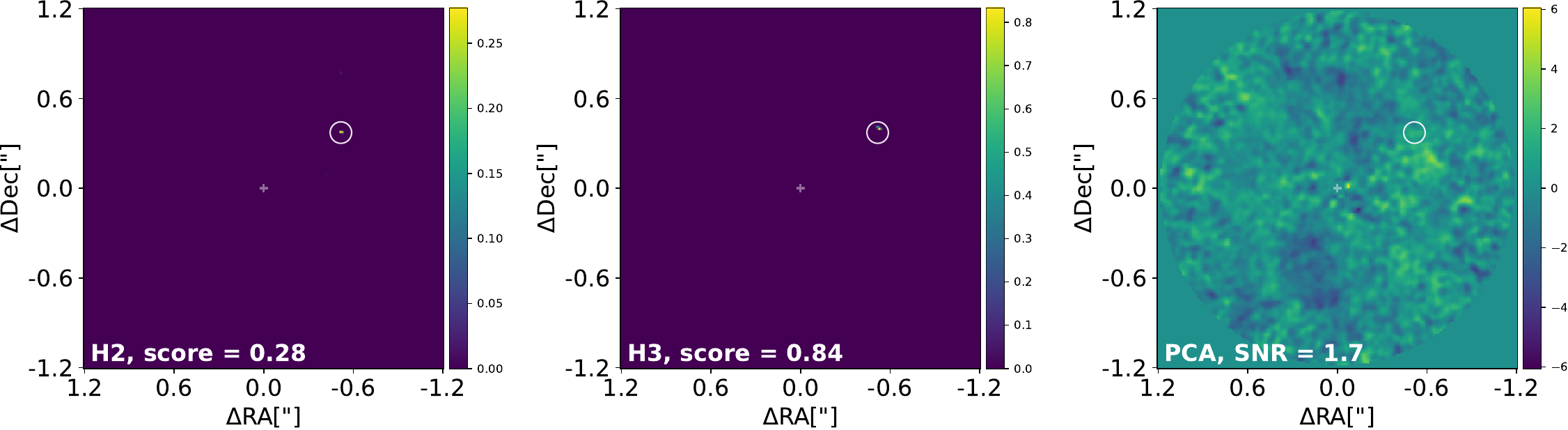}  
    \end{subfigure}
    \begin{subfigure}[b]{\textwidth}
        \centering        \includegraphics[width=0.9\textwidth]{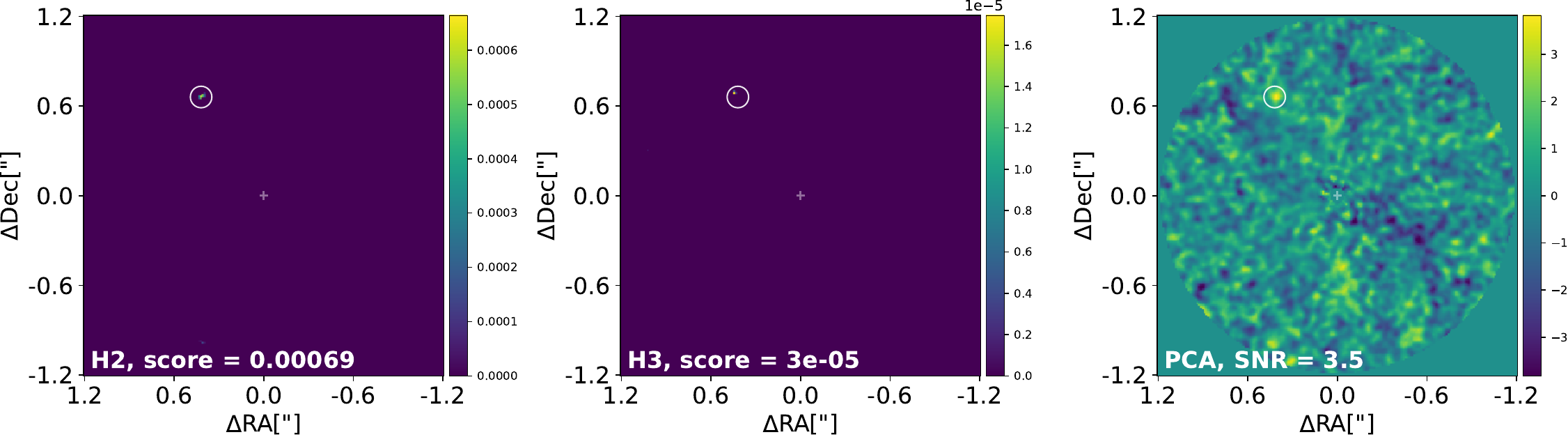}  
    \end{subfigure}
    \begin{subfigure}[b]{\textwidth}
        \centering        \includegraphics[width=0.9\textwidth]{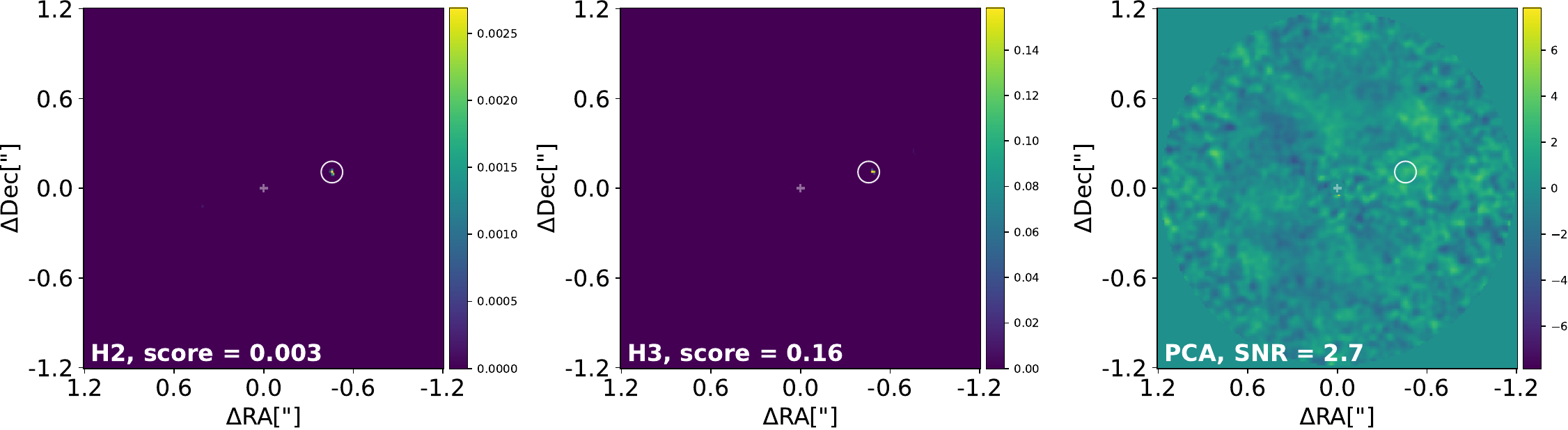}  
    \end{subfigure}

    \caption{Comparison of RSM and PCA S/N maps for several newly identified background stars. Left and middle column panels: RSM maps generated with the APCA–NMF–LOCI–FMKLIP combination in the H2 and H3 filters, along with their corresponding RSM scores. Right column panels: PCA S/N maps computed with five components.}
\end{figure*}

\end{document}